\documentclass[12pt]{article}
\usepackage{epsfig}

\def\dfrac#1#2{{\displaystyle {#1 \over #2}}}

\def\simge{\mathrel{\rlap{\raise 0.511ex \hbox{$>$}}{\lower 0.511ex 
 \hbox{$\sim$}}}}
\def\simle{\mathrel{\rlap{\raise 0.511ex \hbox{$<$}}{\lower 0.511ex 
 \hbox{$\sim$}}}} 
\def\slash#1{\setbox0=\hbox{$#1$}\dimen0=\wd0 \setbox1=\hbox{/} \dimen1=\wd1 
 \ifdim\dimen0>\dimen1 \rlap{\hbox to \dimen0{\hfil/\hfil}} #1 
 \else \rlap{\hbox to \dimen1{\hfil$#1$\hfil}} / \fi}                   
\def\nn{\nonumber}

\newcommand{\ct}{\mathcal}
\newcommand{\beq}{\begin{equation}}
\newcommand{\eeq}{\end{equation}}
\newcommand{\bea}{\begin{eqnarray}}
\newcommand{\eea}{\end{eqnarray}}
\newcommand{\msb}{\overline{\rm{MS}}}
\newcommand{\mev}{\,{\rm MeV}}   
\newcommand{\gev}{\,{\rm GeV}}   
\newcommand{\ri}{{\rm RI/MOM}}
\newcommand{\Oa}{{\cal O}(a)}
\newcommand{\Oaa}{{\cal O}(a^2)}
\newcommand{\Dslash}{\slash D}
\newcommand{\rDslash}{{\overrightarrow{\Dslash}}}
\newcommand{\lDslash}{{\overleftarrow{\Dslash}}}
\newcommand{\dslash}{\slash \partial}
\newcommand{\rdslash}{{\overrightarrow{\dslash}}}
\newcommand{\ldslash}{{\overleftarrow{\dslash}}}
\newcommand{\pslash}{\slash p}
\newcommand{\cp}{{c'\!}}
\newcommand{\cngi}{c_{\scriptstyle{\rm NGI}}}
\newcommand{\Tr}{\hbox{\rm Tr}\,}
\newcommand{\qbar}{{\overline{q}}}

\def\hat{\widehat}

\begin{document}
\thispagestyle{empty}
\begin{flushright}
\begin{tabular}{l}
{\tt DESY 04-007} \\
{\tt FTUV-04-0120} \\
{\tt IFIC/04-02} \\
{\tt LPT Orsay, 04-01}\\
{\tt RM3-TH/04-1} \\
{\tt Roma 1366/04}
\end{tabular}
\end{flushright}

\begin{center}
\vskip 0.6cm
{\Large \bf Renormalization Constants of Quark Operators\\ 
\vskip 0.3cm  for the Non-Perturbatively Improved Wilson Action}
\vskip1.0cm 
{\large\sc D.~Be\'cirevi\'c$^{\,a}$, V.~Gimenez$^{\,b}$, V.~Lubicz$^{\,c}$, 
G.~Martinelli$^{\,d}$, \\ \vskip0.1cm
M.~Papinutto$^{\,e}$ and J.~Reyes$^{\,d}$}\\

\vspace{0.8cm}
{\normalsize {\sl 
$^a$ Laboratoire de Physique Th\'eorique (B\^at.210), Universit\'e de Paris XI,
\\ Centre d'Orsay, 91405 Orsay-Cedex, France.\\ 
\vspace{.25cm}
$^b$ Dep. de F\'isica Te\`orica and IFIC, Univ. de Val\`encia, \\
Dr. Moliner 50, E-46100, Burjassot, Val\`encia, Spain. \\
\vspace{.25cm}
$^c$ Dip. di Fisica, Univ. di Roma Tre and INFN, Sezione di Roma III, \\
Via della Vasca Navale 84, I-00146 Rome, Italy. \\
\vspace{.25cm}
$^d$ Dip. di Fisica, Univ. di Roma ``La Sapienza" and INFN, Sezione di Roma,\\ 
P.le A. Moro 2, I-00185 Rome, Italy. \\
\vspace{.25cm}
$^e$ NIC/DESY Zeuthen, Platanenallee 6, D-15738 Zeuthen, Germany. }}\\
\vskip0.8cm
\end{center}

\begin{abstract}
We present the results of an extensive lattice calculation of the 
renormalization constants of bilinear and four-quark operators for the 
non-perturbatively $\Oa$-improved Wilson action. The results are obtained in the
quenched approximation at four values of the lattice coupling by using the 
non-perturbative RI/MOM renormalization method. Several sources of systematic 
uncertainties, including discretization errors and final volume effects, are 
examined. The contribution of the Goldstone pole, which in some cases may affect
the extrapolation of the re\-nor\-ma\-li\-za\-tion constants to the chiral 
limit, is non-perturbatively subtracted. The scale independent renormalization 
constants of bilinear quark operators have been also computed by using the 
lattice chiral Ward identities approach and compared with those obtained with 
the RI-MOM method. For those renormalization constants the non-perturbative 
estimates of which have been already presented in the literature we find an 
agreement which is typically at the level of 1\%.
\end{abstract}

\vspace*{0.2cm} 
{\small{\tt PACS numbers: 11.15.Ha,\ 12.38.Gc,\ 11.10.Gh.
}}

\newpage

\setcounter{page}{1}
\setcounter{footnote}{0}
\setcounter{equation}{0}

\section{Introduction}
\label{sec:intro}
The increasing availability of computing resources has allowed a reduction of 
the statistical errors in current lattice QCD calculations at the level of few 
percent or less. The corresponding systematic uncertainties, however, are
significantly larger, in some cases even by one order of magnitude. For this 
reason, a major effort in lattice calculations is aimed to reduce and to 
better control the various sources of systematic uncertainties. An important 
ingredient, in this respect, is the use of non-perturbative renormalization 
(NPR) techniques. By avoiding the perturbative expansion in the determination of
the renormalization constants (RCs), these techniques remove the uncertainty 
associated with the truncation of the perturbative series and allow to reduce 
the systematic error involved in the renormalization procedure of the lattice 
matrix elements at a level of accuracy which is comparable or better than the 
present statistical one.

The first NPR technique was proposed almost 20 years ago. Within this approach
the RCs are determined by requiring that the renormalized lattice Green 
functions satisfy the proper chiral Ward identities (WIs)~\cite{boch}. This 
theoretically nice and numerically accurate approach suffers, however, of an 
important limitation: the use of chiral WIs only permits the determination of 
scale independent RCs (and mixing coefficients). After~\cite{boch}, it took 
about 10 years before other two NPR methods, which can be used in principle to 
compute any RC, were developed. They are based on the so called 
$\ri$~\cite{rimom} and Schr\"{o}dinger Functional (SF) schemes~\cite{sf}. In the
$\ri$ scheme the renormalization conditions are imposed, non-perturbatively, on 
quark and gluon Green functions at large external momenta, while in the SF 
approach they are imposed on Green functions computed in a small volume. More 
recently, a new proposal for NPR has been suggested in~\cite{XS}, and it is 
based on the study of lattice correlation functions at short distance in 
$x$-space. The results of the first numerical investigations of this approach 
have been presented in~\cite{lat02_lub,lat03_rey}.\footnote{See also 
ref.~\cite{opelatt} for a related study in the context of the lattice non-linear
$\sigma$-model.}

The non-perturbative $\ri$ renormalization method has been already largely 
applied in lattice calculations to determine the RCs of several classes of 
operators for different versions of the lattice action (see for example 
refs.~\cite{Conti:1997qk}-\cite{Becirevic:2001yh}). In this paper we present the
results of an extensive $\ri$ calculation of the RCs of the complete set of 
bilinear and four-fermion $\Delta F=2$ and $\Delta I=3/2$ operators, for the 
non-perturbatively $\Oa$-improved Wilson action~\cite{alpha} in the quenched
approximation. The results are obtained at four values of the lattice coupling, 
in the range $6.0\le \beta \le 6.45$. Particular attention in the present 
calculation has been dedicated to the evaluation and control of systematic 
uncertainties:
\begin{itemize}
\item {\it Discretization errors} have been evaluated by studying the behaviour 
of the RCs as a function of the lattice coupling, $g(a)$, and of the 
renormalization scale. In the case of bilinear quark operators, since the action
and the operators are non-perturbatively improved at $\Oa$, leading 
discretization effects are of $\Oaa$. These effects have been further reduced to
${\cal O}(g^4a^2)$ by using the results of a recent perturbative calculation of 
the relevant correlation functions~\cite{reyes}. On the other hand, the four
fermion operators considered in this study are not improved, and we are left in 
this case with leading discretization effects of $\Oa$. 
\item {\it Finite volume effects} have been examined by comparing the results
of two independent simulations performed, at the same value of the lattice 
coupling ($\beta=6.0$), on different lattice volumes. 
\item {\it Goldstone pole contributions}: power suppressed contributions coming 
from the Goldstone pole, which may affect the extrapolation to the chiral limit 
of the RCs of the pseudoscalar density and of the four-fermion operators 
coupled to the Goldstone boson, have been non-perturbatively subtracted. 
\end{itemize}
The scale independent RCs of the vector and axial-vector currents, $Z_V$ and 
$Z_A$, and the ratio $Z_P/Z_S$, have also been determined in this study by using
the lattice chiral WI approach, and the results are compared with those 
obtained with the $\ri$ method. We also compare our determinations of these
constants with those obtained by the ALPHA~\cite{alpha_zva,alpha_zsp} and 
LANL~\cite{lanl} Collaborations by using the WI method, and our determination of
$Z_P$ with the one obtained by ALPHA~\cite{alpha_zm} within the SF approach. We 
find an agreement which is typically at the level of 1\%. Since the systematics
involved in these approaches are different, these comparisons provide additional
confidence on the high level of accuracy reached in the implementation of the 
$\ri$ non-perturbative method.

Our final results for the RCs of bilinear quark operators and four-fermion
operators are collected in tables~\ref{table:zeta1}-\ref{tab:PV-}. Preliminary 
results of the present study have been presented at the Lattice 2002
conference~\cite{lat02_lub,lat02_rey}.

The plan of this paper is as follows. In sec.~\ref{sec:details} we give the
details of the numerical simulation and briefly review the non-perturbative 
$\ri$ method used to determine lattice RCs. The various sources of systematic 
errors involved in the calculation are discussed in sec.~\ref{sec:systematic}, 
where we also provide estimates of the corresponding uncertainties. The results
for the RCs of bilinear quark operators obtained with the $\ri$ method are 
summarized in sec.~\ref{sec:res2}. In this section, we also compare these 
results with those obtained by using the WI method, with predictions of one-loop
boosted perturbation theory and with the results of 
refs.~\cite{alpha_zva}-\cite{alpha_zm}. The $\ri$ determination of the RCs of 
the four-fermion $\Delta F=2$ operators is discussed in sec.~\ref{sec:fourf} and
we end the paper by presenting our conclusions. More technical issues concerning
the $\Oa$-improvement of the RCs and the Goldstone pole contributions to the 
Green functions of four-fermion operators are collected in appendix A and B 
respectively. 

\section{Details of the lattice calculation}
\label{sec:details}

The lattice parameters used in this study are summarized in 
table~\ref{tab:details}.
\begin{table}
\centering 
\begin{tabular}{|c|ccccc|}  \hline \hline
{\phantom{\huge{l}}}\raisebox{-.2cm}{\phantom{\Huge{j}}}
$ \beta = 6/g_0^2$ & 6.0 & 6.0 & 6.2 & 6.4 & 6.45    \\ 
{\phantom{\huge{l}}}\raisebox{-.2cm}{\phantom{\Huge{j}}}
$ c_{SW}$ &   1.769 &  1.769 & 1.614 & 1.526  & 1.509 \\ 
{\phantom{\huge{l}}}\raisebox{-.2cm}{\phantom{\Huge{j}}}
$ L^3 \times T $&  $16^3 \times 52$ & $24^3 \times 64$ & 
$24^3 \times 64$  & $32^3 \times 70$& $32^3 \times 70$ \\ 
{\phantom{\huge{l}}}\raisebox{-.2cm}{\phantom{\Huge{j}}}
$ \#\ {\rm conf.}$& 500 & 340 &  200 & 150  & 100\\
{\phantom{\huge{l}}}\raisebox{-.2cm}{\phantom{\Huge{j}}}
$a^{-1}(GeV)$ & 2.00(10) & 2.00(10) & 2.75(14) & 3.63(18) & 3.87(19) \\ 
\hline
{\phantom{\huge{l}}}\raisebox{-.2cm}{\phantom{\Huge{j}}}
$\kappa_1$& 0.1335 & 0.13300 & 0.1339 & 0.1347 & 0.1349   \\ 
{\phantom{\huge{l}}}\raisebox{-.2cm}{\phantom{\Huge{j}}}
$\kappa_2$& 0.1338 & 0.13376 & 0.1344 &  0.1349 &  0.1351  \\ 
{\phantom{\huge{l}}}\raisebox{-.2cm}{\phantom{\Huge{j}}}
$\kappa_3$& 0.1340 & 0.13449 & 0.1349 &  0.1351 &  0.1352  \\ 
{\phantom{\huge{l}}}\raisebox{-.2cm}{\phantom{\Huge{j}}}
$\kappa_4$& 0.1342 & ------ & 0.1352 &  0.1353 &  0.1353  \\
{\phantom{\huge{l}}}\raisebox{-.2cm}{\phantom{\Huge{j}}}
$\kappa_{cr}$& 0.135225(5)& 0.135217(7) & 0.135815(3)& 0.135747(2) 
& 0.135686(2) \\ 
\hline \hline
\end{tabular}
{\caption{\label{tab:details} \sl \small Summary of the lattice details and 
parameters used in this work. We also give the values of the inverse lattice 
spacing and of the critical hopping parameter ($\kappa_{cr}$).}}
\end{table}
We have generated ${\cal O}(1000)$ gauge configurations in the quenched
approximation, by using the non-perturbatively $\Oa$-improved Wilson action. The
improvement coefficient $c_{SW}$ has been determined in ref.~\cite{alpha}, as a 
function of the coupling constant. We have considered four different values of 
the lattice coupling, namely $\beta=6.0$, $6.2$, $6.4$ and $6.45$, corresponding
to inverse lattice spacings varying approximately between 2 and 4 GeV. The
estimates of the inverse lattice spacing given in table~\ref{tab:details}, and 
used in the numerical analysis, have been determined from the study of the 
static quark anti-quark potential~\cite{necco} by setting the scale 
$r_0=0.530(25)$ fm, which corresponds to $a^{-1}(\beta=6)=2.0(1)\gev$. The
uncertainty takes into account the typical spread of the results obtained in the
determination of the lattice spacing in the quenched theory. We emphasize that 
mainly ratios of scales, rather than their absolute values, are relevant for the
present calculation and for these ratios the static quark anti-quark potential 
provides an accurate determination. 

For each value of the coupling, the RCs have been computed at four values of the
light-quark mass and eventually extrapolated to the chiral limit.\footnote{The 
$\ri$ is, by definition, a mass independent renormalization scheme.} The values 
of the Wilson hopping parameter used in the simulations are listed in 
table~\ref{tab:details}, together with the corresponding critical values
determined from the vanishing of the axial WI mass. The Wilson parameters 
correspond to bare quark masses which, in lattice units, are in the range $0.01 
\le a\,m_q \le 0.05$. The same set of gauge configurations and quark propagators
used in this study have been also used in ref.~\cite{qm_spqr}, where additional
details on the simulation can be found. In order to study finite volume effects 
we have considered two independent lattice simulations at $\beta=6.0$, performed
on different lattice sizes. The main purpose of the simulation on the larger 
volume was the study of  $K\to\pi\pi$ decays~\cite{kpp1,kpp2} and this motivates
the different choice of the values of quark masses.

The non-perturbative determination of the RCs with the $\ri$ method is based on
the numerical evaluation, in momentum space, of correlation functions of the 
relevant operators between external quark and/or gluon states. In the case of 
the bilinear quark operators $O_\Gamma = \qbar \Gamma q$, where $\Gamma = S,P,V,
A,T$ stands respectively for $I, \gamma_5, \gamma_\mu, \gamma_\mu\gamma_5, 
\sigma_{\mu\nu}$, the relevant Green function is
\beq
{G}_\Gamma(p,p') = \int d^4x \, d^4y \, e^{-i p\cdot x + i p'\cdot y} \, 
\langle \hat q(x) O_\Gamma(0) \hat\qbar(y) \rangle \, ,
\label{eq:gg}
\eeq
where $\hat q$ and $\hat\qbar$ are renormalized quark fields. By using ${G}_
\Gamma(p,p')$ and the renormalized quark propagator,
\beq
\hat{S}(p) = \int d^4x \, e^{-i p\cdot x} \, \langle \hat q(x) \hat\qbar(0) 
\rangle \, ,
\eeq
one evaluates the forward amputated Green function
\beq
\Lambda_\Gamma(p) = \hat{S}(p)^{-1} {G}_\Gamma(p,p) \hat{S}(p)^{-1} \, .
\label{eq:ll}
\eeq
The $\ri$ renormalization method consists in imposing that the forward amputated
Green function, computed in the chiral limit in a fixed gauge and at a given 
(large) scale $p^2=\mu^2$, is equal to its tree-level value. In this study we
work in the Landau gauge, but different choices of generic covariant gauges have
been also considered in the literature~\cite{gauge}. In practice, the 
renormalization condition is implemented by requiring
\beq
Z_\Gamma \, \Gamma_\Gamma(p) \vert_{p^2=\mu^2} \equiv
Z_\Gamma \, \Tr [\Lambda_\Gamma (p)\, P_\Gamma] \vert_{p^2=\mu^2} =1
\label{eq:rimom}
\eeq
in the chiral limit, where $P_\Gamma$ is a Dirac projector satisfying ${\rm Tr}
\, [\Gamma \, P_\Gamma] =1$. The RC of the quark field, which enters the
definition of the renormalized quark propagator, is obtained by imposing the 
condition
\beq
\frac{i}{12}\,\Tr\left[\frac{\slash p\,\hat{S}(p)^{-1}}{p^2}\right]_{p^2=\mu^2}=
Z_q \,\frac{i}{12}\,\Tr\left[\frac{\slash p\, S(p)^{-1}}{p^2}\right]_{p^2=\mu^2}
=1 \,,
\label{eq:zqri}
\eeq 
in the chiral limit. In the numerical calculation, throughout this study, we 
adopt the definitions
\beq
p^2 \equiv \frac{1}{a^2}\, \sum_{\nu=1}^4 \sin^2(a\, p_\nu) \qquad , \qquad
\slash p \equiv \frac{1}{a}\, \sum_{\nu=1}^4 \gamma_\nu \sin(a\, p_\nu)
\eeq 
and, in order to minimize discretization effects, we select momenta with 
components $p_\nu=(2\pi/L_\nu)\, n_\nu$ in the intervals $n_\nu=([0,2],[0,2],
[0,2],[0,3])$ and $n_\nu=([2,3],[2,3],[2,3],[4,7])$ ($L_\nu$ is the lattice size
in the direction $\nu$).
  
The renormalization scale $\mu$, introduced in eqs.~(\ref{eq:rimom}) and 
(\ref{eq:zqri}), must be much larger than $\Lambda_{QCD}$, in order to be able 
to connect the $\ri$ renormalization scheme to other schemes by using 
perturbation theory, and much smaller than the inverse lattice spacing, to avoid
large discretization errors:
\beq 
\Lambda_{QCD} \ll \mu \ll \pi/a \, .
\eeq

An important point to be mentioned is that the RCs of bilinear quark operators,
obtained with the $\ri$ technique in the chiral limit and at sufficiently large 
values of the external momentum, are automatically improved at $\Oa$. The reason
is related to chiral symmetry and to the fact that $\Oa$ on-shell improvement 
corrections, proportional to the coefficients $c_V$, $c_A$ and $c_T$, vanish for
Green functions computed between forward external states, at zero momentum 
transfer. An explicit proof of this statement is given in appendix A. Being the 
improvement coefficients known with some uncertainty, the possibility of 
neglecting them in the calculation improves the accuracy of the determination of
these RCs. In contrast, on-shell $\Oa$-improvement of the operators is required 
when the RCs are computed by using the WI approach. In this case, in order to 
improve the vector and axial-vector current operators, we use the values of the 
coefficients $c_V$ and $c_A$ determined non-perturbatively in 
refs.~\cite{alpha,lanl}.

\section{Systematic errors}
\label{sec:systematic}
In this section we discuss the $\ri$ non-perturbative evaluation of the RCs of
bilinear quark operators by examining in details the various sources of 
systematic errors and providing estimates of the related uncertainties.

\subsection{Renormalization scale dependence and discretization effects}
\label{sec:syst-oa}
\begin{figure}
\hspace*{-0.6cm}
\begin{tabular}{cc}
\epsfxsize8.0cm\epsffile{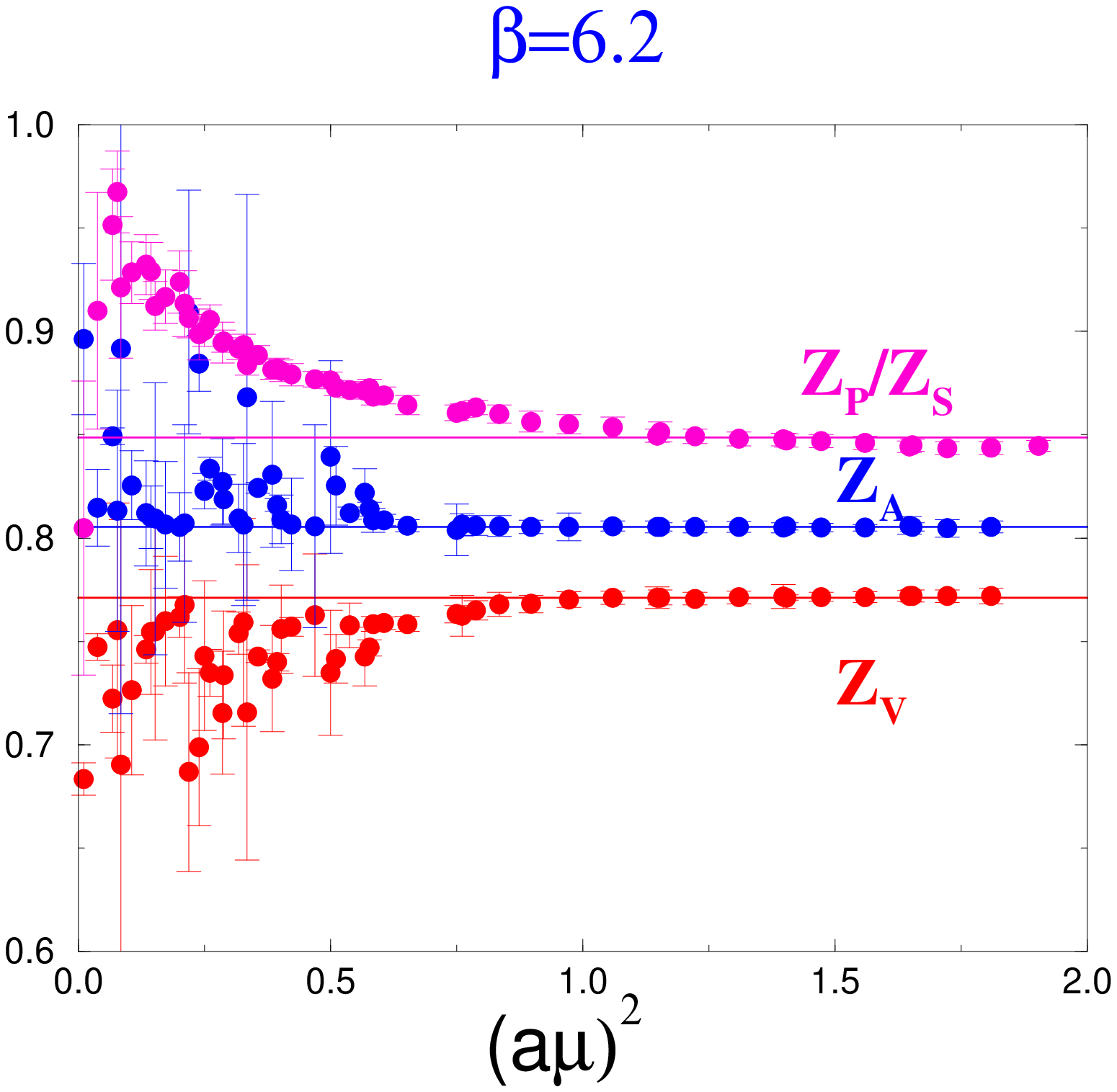} &
\hspace*{-0.5cm}
\epsfxsize8.0cm\epsffile{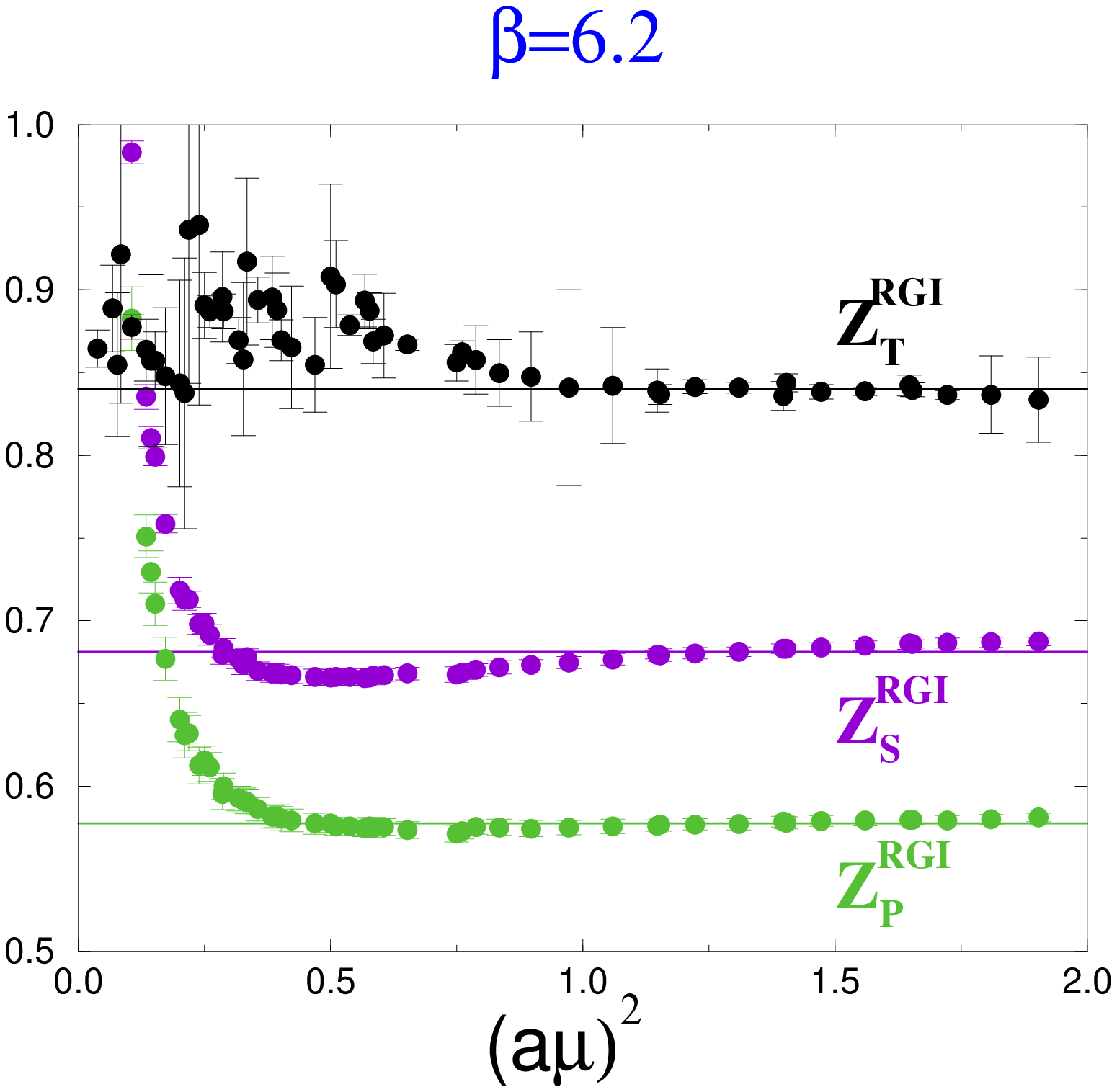} \\
\end{tabular}
\caption{\label{fig:zall62}{\sl \small The scale independent RCs $Z_V$, $Z_A$ 
and the ratio $Z_P/Z_S$ (left) and the RGI combinations $Z_P^{RGI}$, $Z_S^{RGI}$
and $Z_T^{RGI}$ (right) are shown as a function of the renormalization scale, 
at $\beta=6.2$. The solid lines represent the results of a constant fit to the 
data.}}
\end{figure}

The results for the scale independent RCs $Z_V$ and $Z_A$, and for the ratio 
$Z_P/Z_S$, obtained with the $\ri$ method are presented in the left plot of
fig.~\ref{fig:zall62} as a function of the renormalization scale. We show, as an
example, the results obtained at $\beta=6.2$.
The good quality of the plateau indicates that $\Oaa$ discretization effects are
rather small, at the level of the statistical errors, even in the region $a\mu 
\simge 1$ considered here. 

For the scale dependent bilinear operators, ${\cal O}=S,P,T$, we show in the 
right plot of fig.~\ref{fig:zall62} the renormalization group invariant (RGI) 
combinations
\beq
Z_{\cal O}^{RGI} = Z_{\cal O}(\mu)/C_{\cal O}(\mu) \, ,
\eeq
where the evolution function $C_{\cal O}(\mu)=\exp\left[\int^{\,\alpha(\mu)} d
\alpha \,\gamma_{\cal O}(\alpha)/\beta(\alpha)\right]$, with $\beta(\alpha)$ and
$\gamma_{\cal O}(\alpha)$ the beta function and the anomalous dimension of the
relevant operator, is introduced in order to explicitly cancel, at a given order
in perturbation theory, the scale dependence of the RCs. In the $\ri$ scheme, 
these functions are known at the N$^2$LO for $Z_T$~\cite{gracey} and at the 
N$^3$LO for $Z_S$ and $Z_P$~\cite{chetyr}. In their numerical evaluation, we use
the determination of the strong coupling constant obtained, in the quenched
theory, from $\Lambda_{\msb}^{\rm (n_f=0)}=225(20)\mev$~\cite{alpha_zm}.
From the quality of the plateau shown in the right plot of
fig.~\ref{fig:zall62}, discretization effects appear to be very small, even at 
(unexpectedly) large values of $a\mu$.\footnote{The Goldstone pole contributions
to the ratio $Z_P/Z_S$ and to the combination $Z_P^{RGI}$ shown in 
fig.~\ref{fig:zall62} have been non-perturbatively subtracted as discussed in 
sec.~\ref{sec:syst-goldstone}.}

\subsection{Perturbative correction of ${\cal O}(g^2a^2)$ discretization 
effects}
\label{sec:syst-pert}
Any deviation from the predicted constant behaviour at large momenta, as 
observed in fig.~\ref{fig:zall62} and in analogous results obtained at other 
values of $\beta$, signals the presence of either $\Oaa$ discretization effects
or of higher-order perturbative corrections not included in the evaluation of 
the evolution function $C_{\cal O}(\mu)$. These deviations are found to be 
larger on the coarser lattice, corresponding in our case to $\beta=6.0$. One can
also notice in fig.~\ref{fig:zall62} that the quality of the plateau is worse in
the case of the RC of the scalar density $Z_S$. As discussed below, we interpret
this to be due to the presence of larger discretization errors.

The leading ${\cal O}(g^2a^2)$ discretization errors in the lattice expressions 
of the amputated Green functions $\Lambda_\Gamma(p)$ of eq.~(\ref{eq:ll}) have 
been recently computed by using lattice perturbation theory~\cite{reyes}. In 
order to further investigate the effect of discretization errors in the 
calculation of the RCs, we use the results of \cite{reyes} and subtract the 
perturbative ${\cal O}(g^2a^2)$ contributions from the amputated Green functions
$\Lambda_\Gamma(p)$ computed non-perturbatively in the numerical simulation.

\begin{figure}
\hspace*{-0.6cm}
\begin{tabular}{cc}
\epsfxsize8.0cm\epsffile{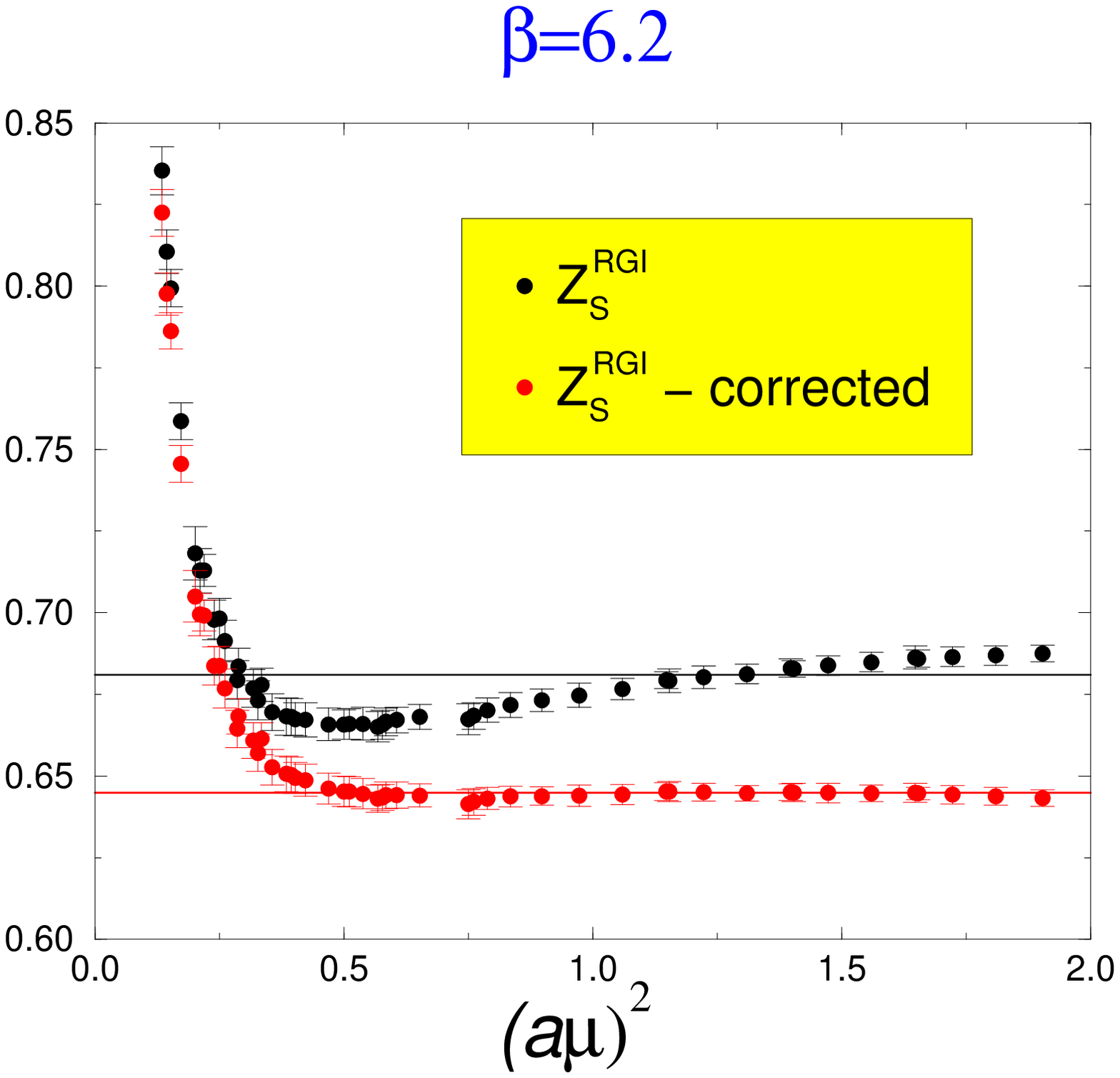} &
\hspace*{-0.5cm}
\epsfxsize8.0cm\epsffile{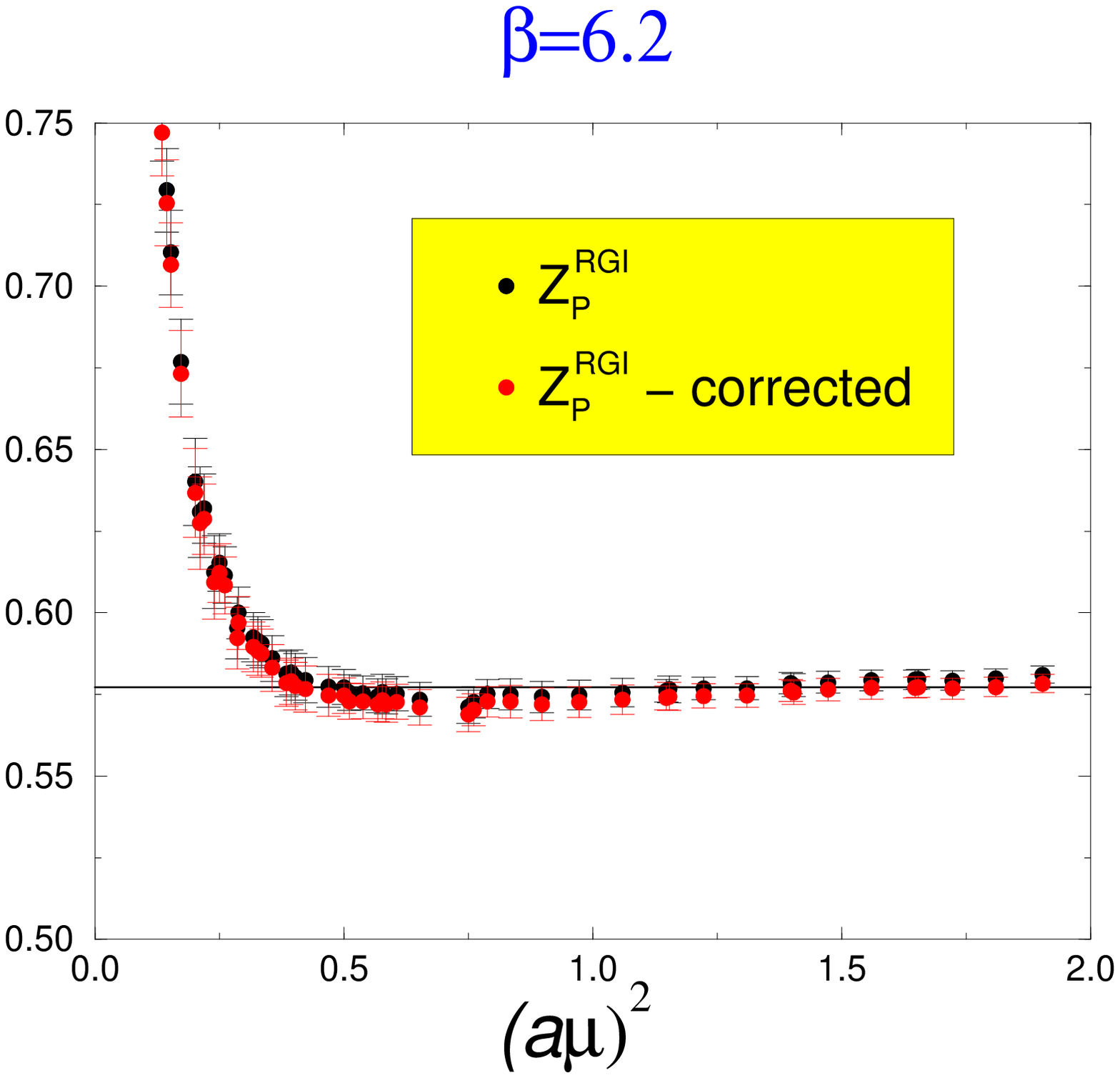} \\
\end{tabular}
\caption{\label{fig:zzoa}{\sl \small The RGI combinations $Z_S^{RGI}$ (left) and
$Z_P^{RGI}$ (right) as obtained with and without the subtraction of the 
${\cal O}(g^2a^2)$ contributions computed in perturbation theory.}}
\end{figure}

The most sensitive to the subtraction of the ${\cal O}(g^2a^2)$ terms is the 
constant $Z_S$. The correction decreases the final estimate of this RC by 
approximately 5\% and we find that after the subtraction the quality of the 
plateau in the RGI combination is significantly improved. The result is shown in
the left plot of fig.~\ref{fig:zzoa}, at $\beta=6.2$. On the other hand, the 
effect of the ${\cal O}(g^2a^2)$ correction is found to be negligible for $Z_P$,
as shown in the right panel of fig.~\ref{fig:zzoa}. In the case of the RCs $Z_T$
the correction amounts to approximately 3\% and for $Z_V$ and $Z_A$ to 
approximately 1\%.

All the results for the RCs of bilinear quark operators discussed in the 
following will be those with the terms of ${\cal O}(g^2a^2)$ subtracted away. 
After this correction, the systematic error due to residual $\Oaa$ effects is 
evaluated from the quality of the plateau of the RGI combinations as a function
of the renormalization scale. This error will be included in the final 
evaluation of the RCs presented in sec.~\ref{sec:res2}

\subsection{Comparison of results at different values of the coupling}
\label{sec:syst-allbeta}
\begin{figure}
\hspace*{-0.6cm}
\begin{tabular}{cc}
\epsfxsize8.0cm\epsffile{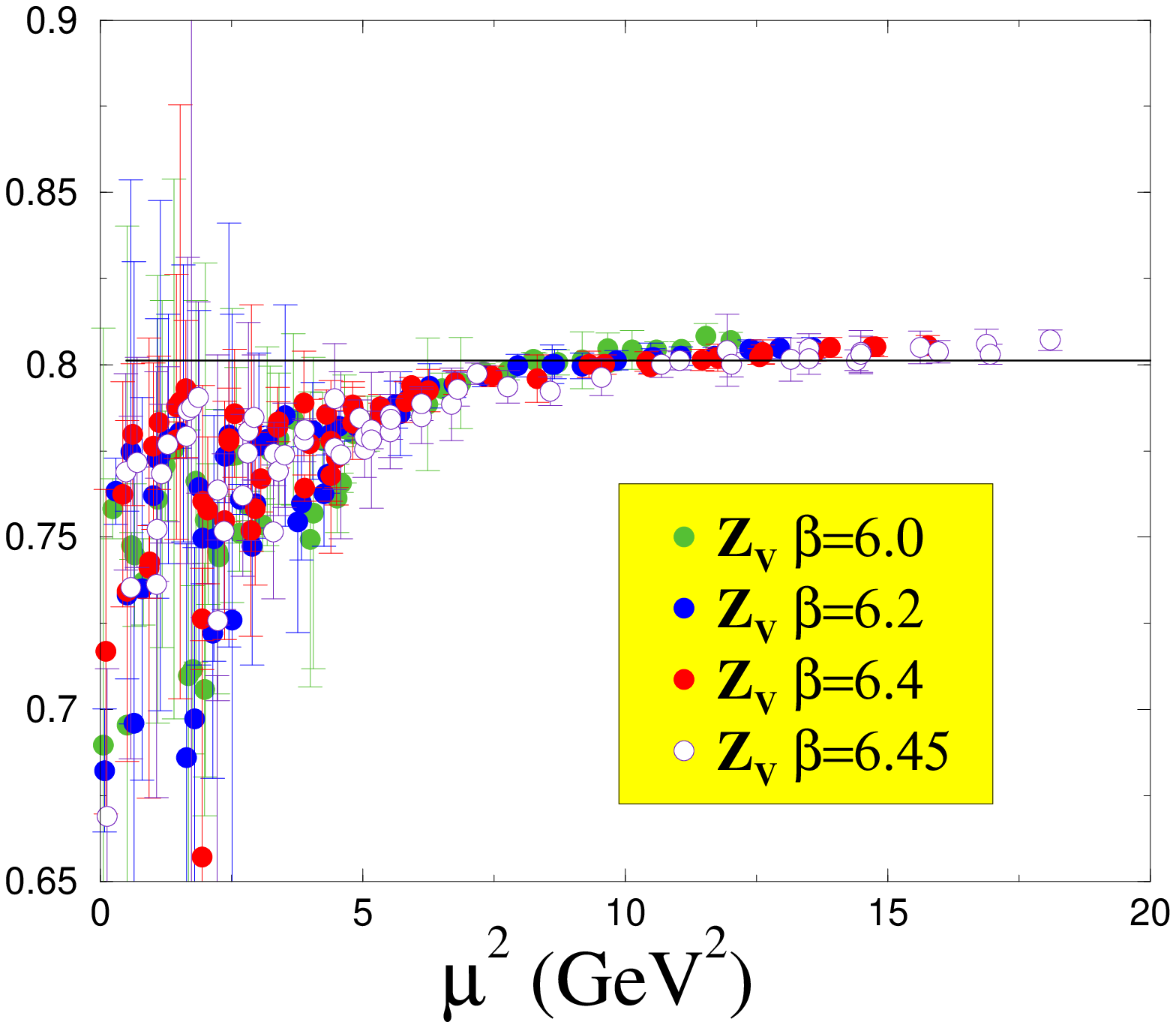} &
\hspace*{-0.5cm}
\epsfxsize8.0cm\epsffile{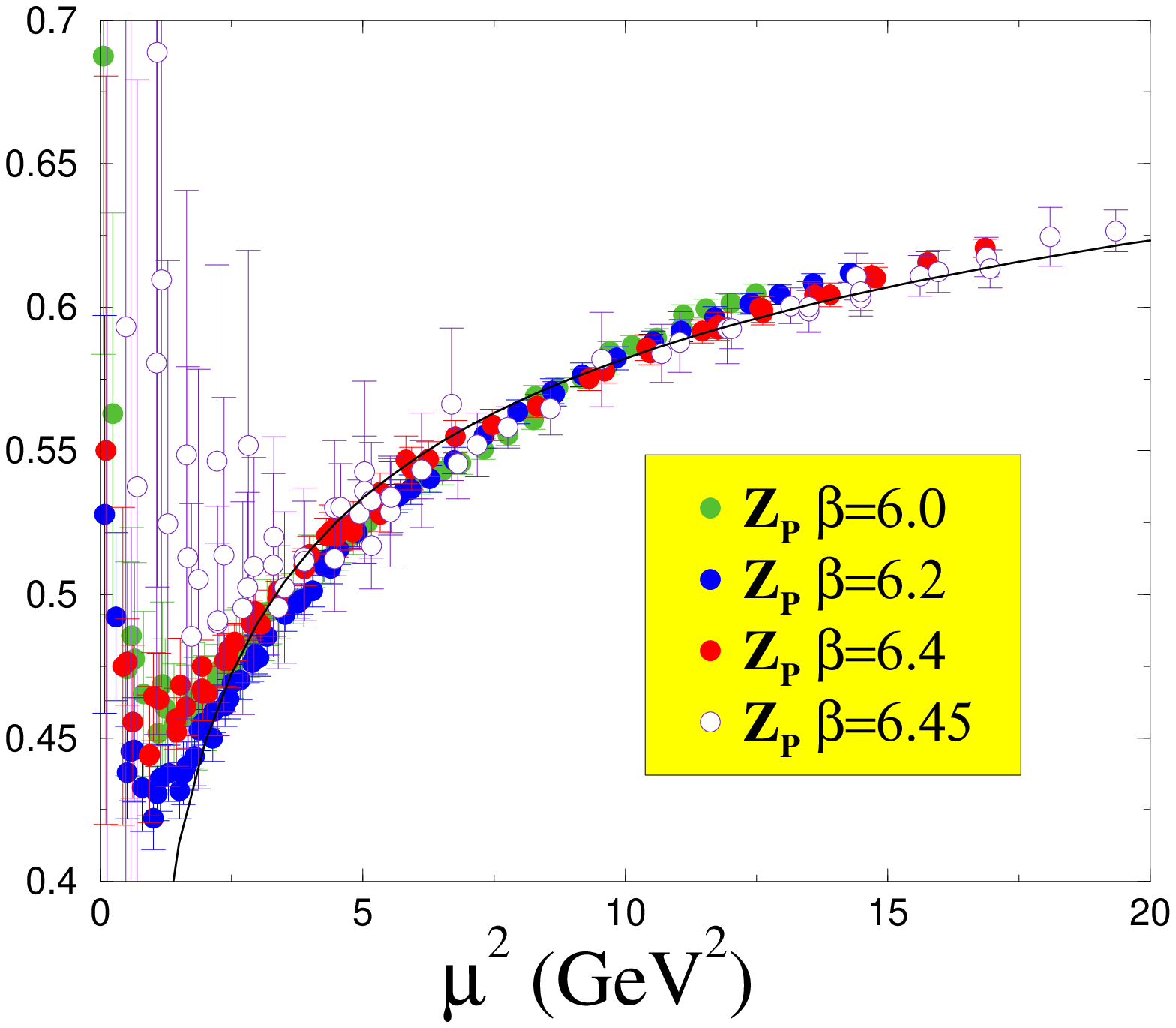} \\
\end{tabular}
\caption{\label{fig:zrisc}{\sl \small Values of $Z_V$ (left) and $Z_P$ (right) 
as obtained at the four values of the lattice coupling as a function of the 
renormalization scale. The results have been rescaled by the factor $R(a,\bar 
a)$ defined in eq.~(\ref{eq:erre}). The solid lines represent the scale 
dependence predicted by the relevant anomalous dimensions ($\gamma=0$ in the 
case of $Z_V$).}}
\end{figure}
Another possibility to study discretization effects is based on the comparison 
of the RCs computed at different values of the lattice spacing. Though the RCs 
are obviously functions of the lattice spacing (the UV cutoff in the lattice 
regularization), their dependence on the renormalization scale should be 
universal (i.e. independent of $a$), and only fixed by the anomalous dimension 
of the corresponding operators. Therefore, the dependence on the renormalization
scale should cancel in the ratio
\beq
\label{eq:erre}
R(a,\bar a) \equiv Z(\bar a,\mu)/Z(a,\mu)\, ,
\eeq
up to discretization effects. In fig.~\ref{fig:zrisc} we show as an example the 
results for the RCs $Z_V$ and $Z_P$ obtained, at the four values of $\beta$, as 
a function of the renormalization scale. Each RC, $Z(a,\mu)$, has been rescaled
by the factor $R(a,\bar a)$ of eq.~(\ref{eq:erre}), where we have chosen as a 
reference scale $\bar a$ the value of the lattice spacing at $\beta=6.4$. From 
fig.~\ref{fig:zrisc}, we observe that the results obtained at the different 
values of $\beta$ all lies, for large values of the renormalization scale, on 
the same universal curve, thus confirming that discretization effects are well 
under control within the statistical errors. The figure also shows that the 
renormalization scale dependence of the RCs is in very good agreement with the 
one predicted by the anomalous dimensions of the relevant operators. On the
other hand, we note that at small values of the renormalization scale the 
results obtained at different $\beta$s show some disagreement, and that the 
dependence on the renormalization scale differs from the one predicted in 
perturbation theory. These discrepancies may be due to finite volume effects,
higher order perturbative corrections and also to power suppressed 
non-perturbative contributions which might not be negligible in the region of 
small momenta. For this reason, the results obtained at small momenta are not
considered in our final determination of the RCs.

\subsection{Subtraction of the Goldstone pole}
\label{sec:syst-goldstone}
The validity of the $\ri$ approach relies on the fact that non-perturbative 
contributions to the Green functions vanish asymptotically at large $p^2$. Among
the bilinear quark operators, however, specific care must be taken in the study 
of the pseudoscalar Green function $\Gamma_P$ since in this case, due to the 
coupling with the Goldstone boson, the leading power suppressed contribution is 
divergent in the chiral limit~\cite{rimom,alain1,alain2}. The operator product 
expansion (OPE) of $\Gamma_P$ reads in fact:
\beq
\Gamma_P(p^2,m) \simeq c_1(p^2,m) + c_2(p^2,m) \, \frac{\langle \qbar q 
\rangle} {m\, p^2}  + {\cal O}(1/p^4) \, ,
\label{eq:gammap}
\eeq
where $m$ is the quark mass, $\langle \qbar q \rangle$ is the quark condensate
in the chiral limit and $c_{1,2}$ are Wilson coefficients. In practice, since we
work in a region of light quark masses ($m^2/p^2\ll 1$) varying in a short 
interval, the mass dependence of the Wilson coefficients can be neglected. We 
extract the RC $Z_P$ from the short-distance behaviour of $\Gamma_P$ at large 
$p^2$, represented in eq.~(\ref{eq:gammap}) by the coefficient $c_1$. 
In order to determine the contribution of the Wilson coefficient $c_1$ we have 
then followed two different approaches.

\begin{figure}
\begin{center}
\epsfxsize8.0cm
\epsffile{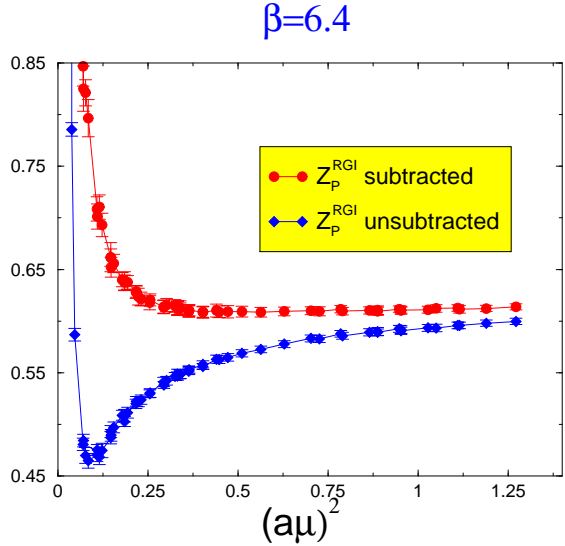} 
\caption{\label{fig:zpnosub}{\sl \small The RGI combination $Z_P^{RGI}$ as 
obtained with and without the non-perturbative subtraction of the Goldstone 
pole, at $\beta=6.4$.}}
\end{center}
\end{figure}

In the first one the Goldstone pole contribution proportional to $c_2$ has 
been subtracted by constructing the combinations~\cite{gv}:
\beq
\label{eq:gpsub}
\Gamma_P^{\rm SUB}(p^2,m_1,m_2) = \frac{m_1\, \Gamma_P(p^2,m_1) - m_2\, 
\Gamma_P(p^2,m_2)}{m_1-m_2} \,,
\eeq
where $m_1$ and $m_2$ are non degenerate quark masses.\footnote{In this analysis
we used the vector WI definition of the quark mass, $m_q=\frac{1}{2}\left(
1/\kappa_q-1/\kappa_{cr}\right)$. We also verified that using the axial WI 
definition leads to practically indistinguishable results.} The effect of the 
subtraction on the RGI combination $Z_P^{RGI}$ is shown in 
fig.~\ref{fig:zpnosub}. 
Though both the unsubtracted and the subtracted determinations of $Z_P^{RGI}$ 
converge to approximately the same value at large $p^2$, a clear plateau is 
never observed in the first case, at least not in the region of momenta explored
in this study. For this reason, we have derived our first determination of the 
RC $Z_P$ by using the subtracted Green function $\Gamma_P^{\rm SUB}$ defined 
in eq.~(\ref{eq:gpsub}).

In the second approach we first study the dependence of $\Gamma_P$ on the quark
mass~\cite{alain1}. Guided by the OPE of eq.~(\ref{eq:gammap}), we have
performed a fit of $\Gamma_P$ to the form\footnote{Since the vertex Green 
function has been computed by inserting operators with both degenerate and 
non-degenerate quark masses, the 3-parameter fit in eq.~(\ref{eq:fit1m}) is 
performed on a set of 10 data points.}
\beq
\label{eq:fit1m}
\Gamma_P(p^2,m) = A(p^2) + \frac{B(p^2)}{m} + C(p^2)\,m \,.
\eeq
The singular Goldstone pole contribution to $\Gamma_P$ is represented in this 
expression by the term proportional to $B(p^2)$. The coefficient $A(p^2)$ 
provides instead, at large $p^2$, the short-distance asymptotic behaviour of the
Green function in the chiral limit. From this coefficient we have therefore 
obtained our second determination of $Z_P$.

We find that the RC determined with the two approaches, namely either from the
subtracted Green function $\Gamma_P^{\rm SUB}$ in eq.~(\ref{eq:gpsub}) or from
the coefficient $A(p^2)$ of eq.~(\ref{eq:fit1m}), are in perfect agreement, the
difference being less than 1\% at all values of the coupling $\beta$.

In order to further investigate the Goldstone pole contribution to the 
pseudoscalar Green function, we have studied the momentum dependence of the
coefficient $B(p^2)$ by performing a fit to the form
\beq
\label{eq:fit1p}
B(p^2) = \alpha + \frac{\beta}{p^2} + \frac{\gamma}{p^4} \,.
\eeq
We find that in the region of relatively large external momenta ($p\simge 2$ 
GeV) this fit provides a good description of the lattice data. The values of the
coefficient $\alpha$, which may be only generated by pure lattice artifacts, are 
found to be consistent with zero at all values of the lattice coupling. From the
results for the coefficient $\beta$ we have derived an estimate of the quark 
condensate, which turns out to be in agreement with other lattice determinations
of the same quantity based on different 
approaches~\cite{Giusti:1998wy}-\cite{Chiu:2003iw}. The details of the 
extraction of the quark condensate from the pseudoscalar Green function are 
presented in a separate publication~\cite{qq_spqr}.

\subsection{Chiral extrapolations}
\label{sec:syst-chiral}
The renormalization condition in eq.~(\ref{eq:rimom}), which defines the $\ri$
renormalization scheme, must be implemented in the chiral limit. This ensures
that $\ri$ is a {\em mass independent} renormalization scheme. Since 
the non-perturbative determinations of the RCs are obtained in the calculation 
at non vanishing values of the quark masses, a final extrapolation of the 
results to the chiral limit must be eventually performed.

\begin{figure}
\hspace*{-0.6cm}
\begin{tabular}{cc}
\epsfxsize8.0cm\epsffile{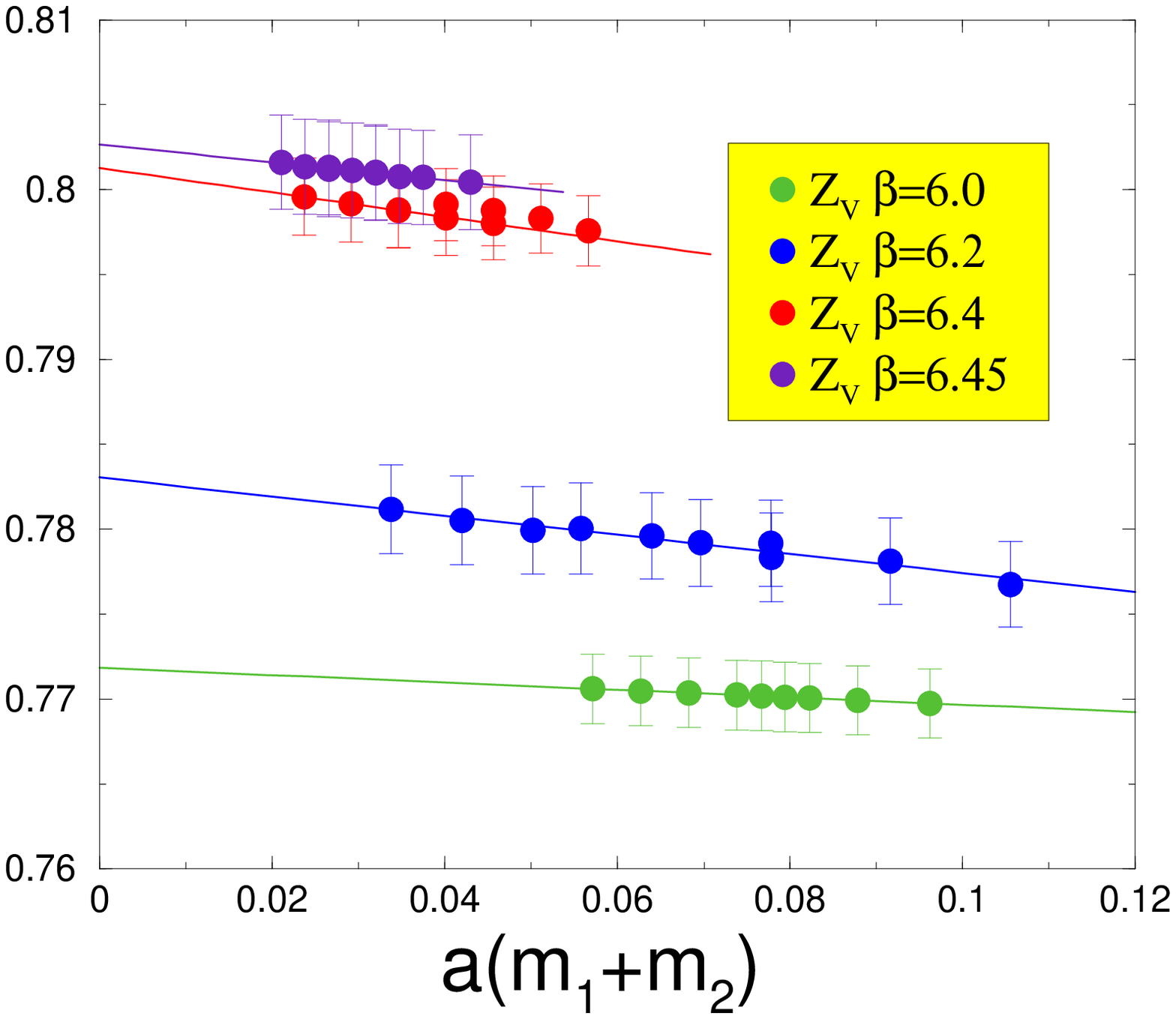} &
\hspace*{-0.5cm}
\epsfxsize8.0cm\epsffile{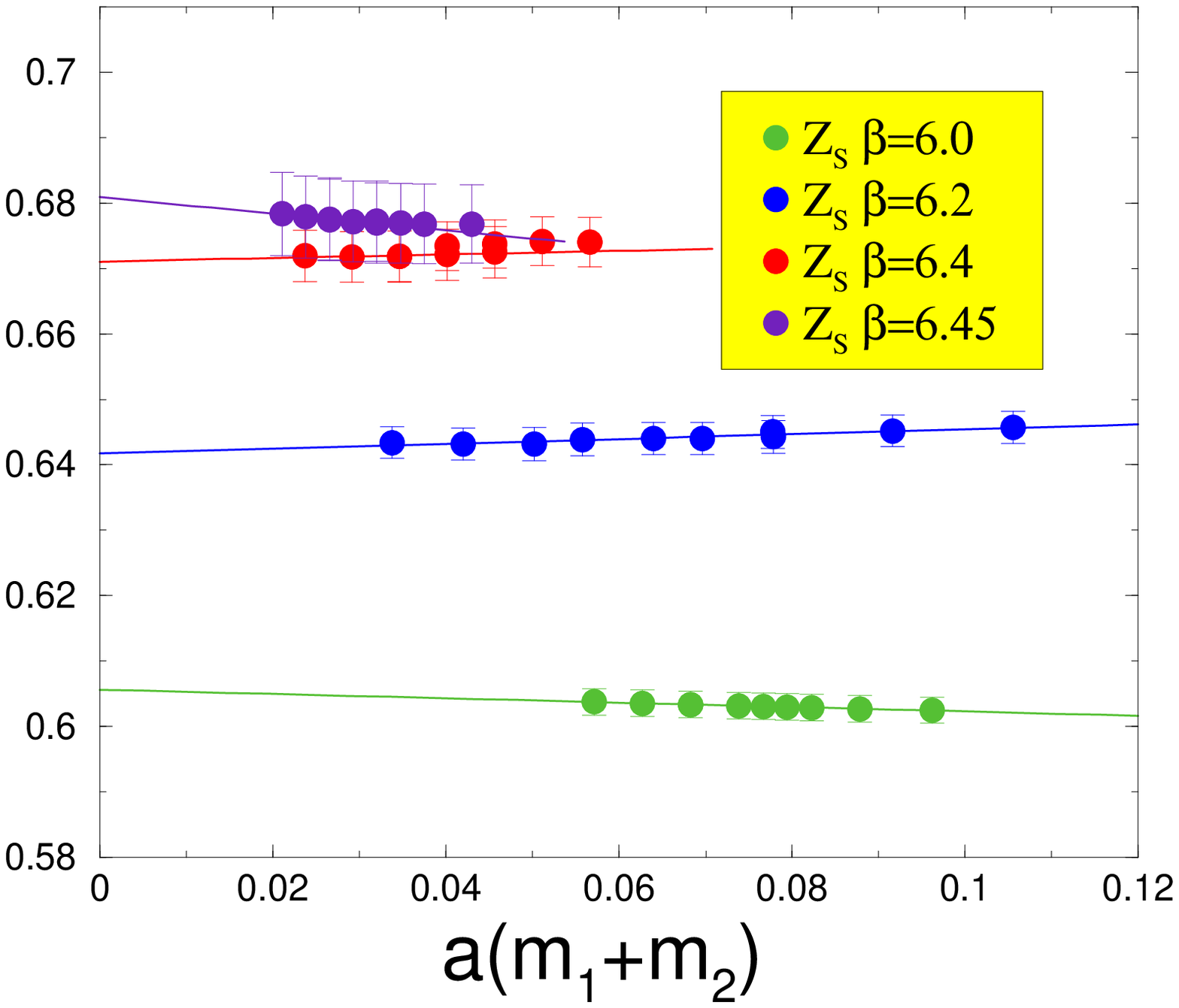} \\
\end{tabular}
\caption{\label{fig:zextra}{\sl \small Values of $Z_V$ (left) and $Z_S$ (right) 
as obtained at the four values of the lattice coupling as a function of the 
bare quark masses. The solid lines represent the results of a linear
extrapolation to the chiral limit.}}
\end{figure}

After the contribution of the Goldstone pole has been non-perturbatively 
subtracted from the pseudoscalar correlator, all Green functions of bilinear 
quark operators are expected to be smooth functions of the quark masses. In the 
region of masses considered in this study, this mass dependence is found to be 
rather weak and consistent with a simple linear behaviour. For this reason, the 
extrapolation to the chiral limit is easily performed. For illustrative
purposes, we show in fig.~\ref{fig:zextra} the chiral extrapolation of the RCs 
$Z_V$ and $Z_S$ at the four values of the lattice coupling.
The smooth dependence on the quark masses suggests that the systematic error
introduced by the chiral extrapolation is smaller than the statistical error and
can be therefore safely neglected.

\begin{figure}
\hspace*{-0.6cm}
\begin{tabular}{cc}
\epsfxsize8.0cm\epsffile{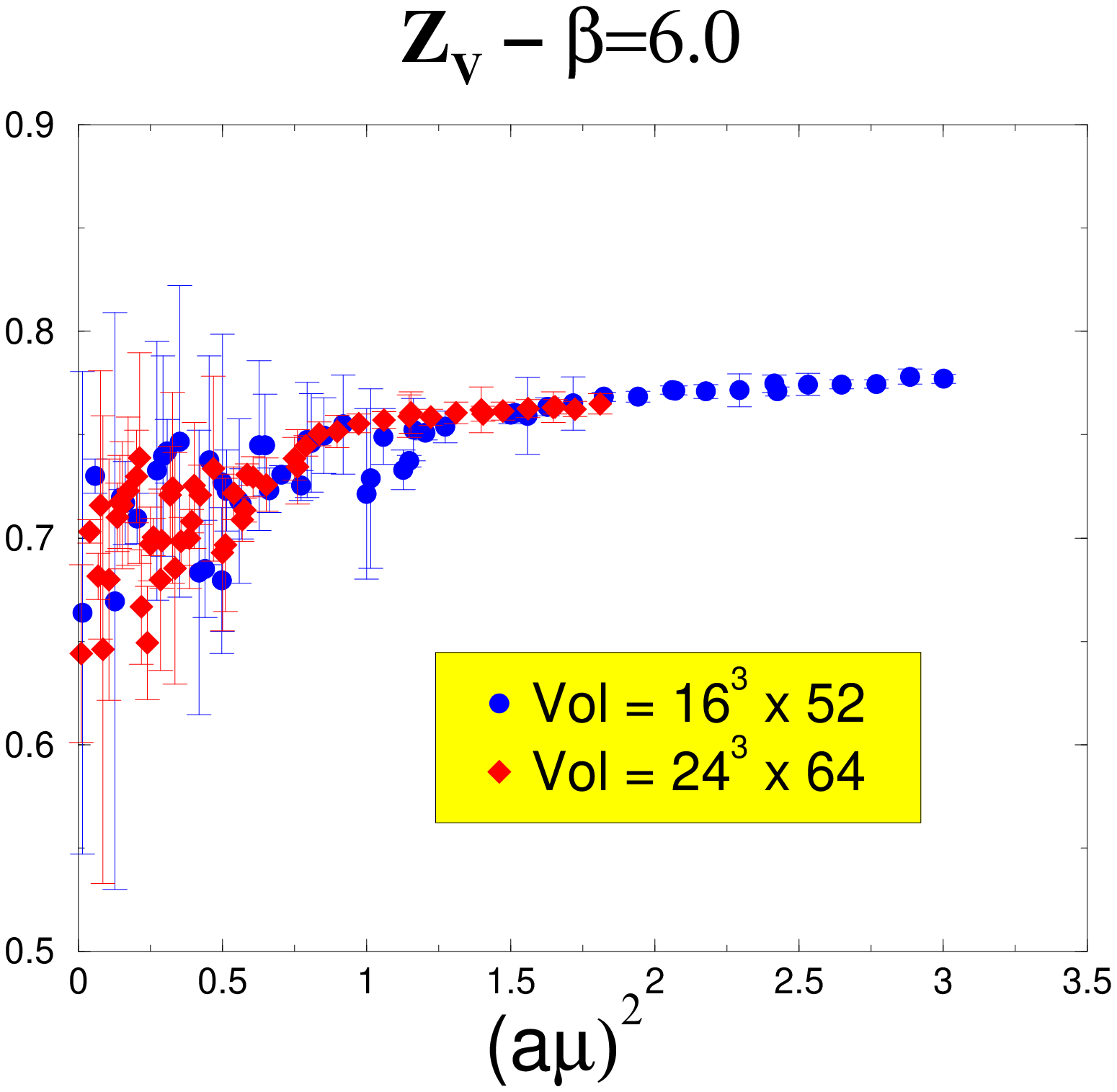} &
\hspace*{-0.5cm}
\epsfxsize8.0cm\epsffile{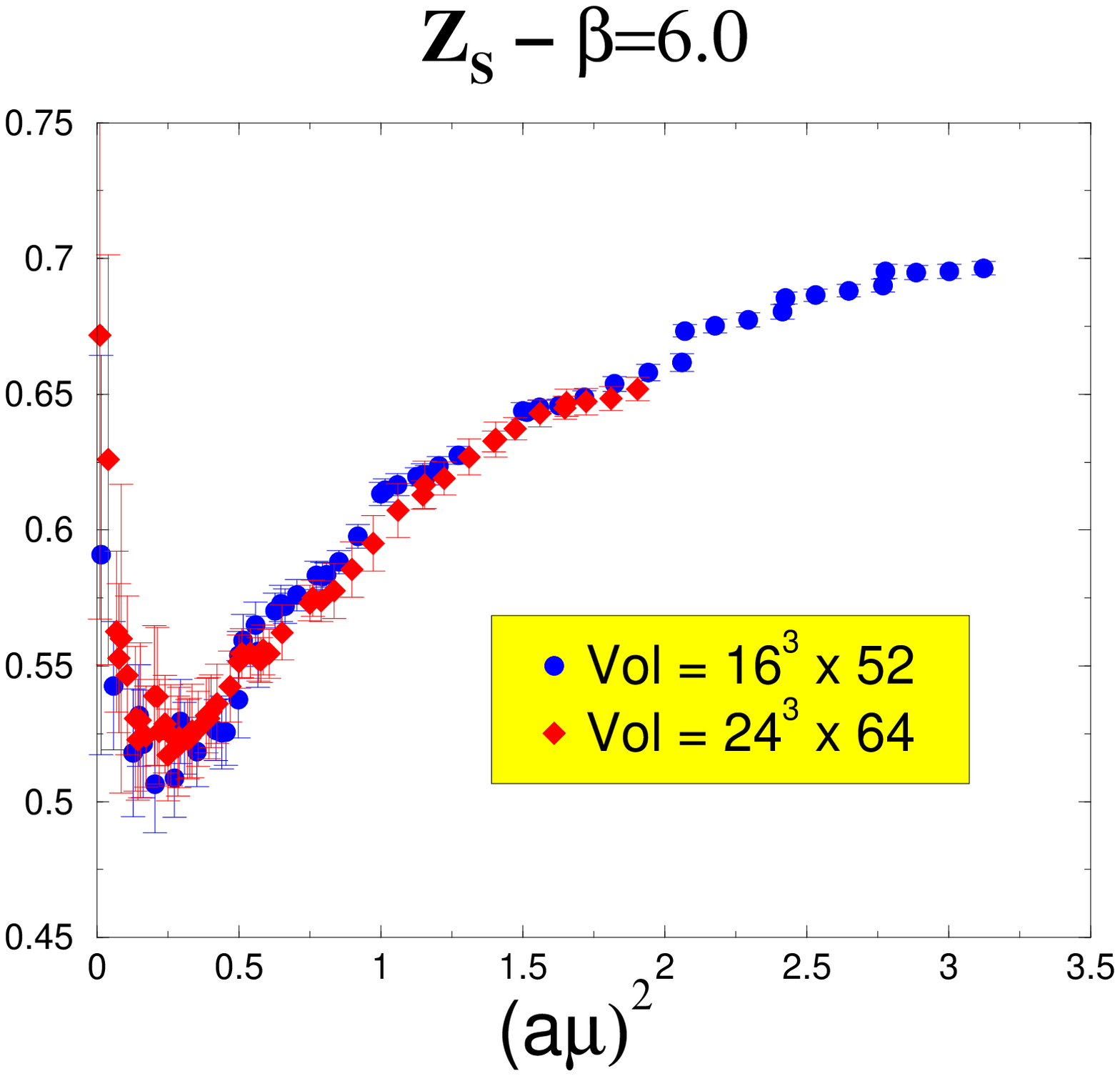} \\
\end{tabular}
\caption{\label{fig:z2vol}{\sl \small Comparison of $Z_V$ (left) and $Z_S$ 
(right) at $\beta=6.0$ as obtained on lattices of different sizes as a function 
of the renormalization scale.}}
\end{figure}
\subsection{Finite volume effects}
\label{sec:syst-vol}
In order to study finite size effects, two independent simulations on different 
lattice sizes at $\beta=6.0$ have been considered. The smallest size corresponds
approximately to the same physical volume used at the other values of $\beta$. 
From this analysis we find that the differences between the values of the RCs
obtained at the same value of the renormalization scales from lattices of 
different sizes are smaller than the statistical errors. For this reason, we
have not included an additional systematic error due to final volume effects 
in our final determination of the RCs at all values of the lattice coupling. An 
example of this comparison, for the RCs $Z_V$ and $Z_S$, is shown in 
fig.~\ref{fig:z2vol}.

\section{Results for the RCs of bilinear quark operators}
\label{sec:res2}
In this section we present our final results for the RCs of bilinear quark
operators as obtained with the $\ri$ method. We then compare these results with 
those obtained by using the WI approach (for the scale independent constants),
one-loop boosted perturbation theory and with other results presented in the
literature.

\subsection{RI/MOM results}
Our final estimates of the RCs of bilinear quark operators obtained with the
$\ri$ method are derived by fitting to a constant the RGI combinations shown in 
fig.~\ref{fig:zall62} (and similarly at the other values of $\beta$) after 
having corrected the ${\cal O}(g^2a^2)$ discretization errors, as discussed in 
sec.~\ref{sec:syst-pert}, and performed the extrapolation to the chiral limit. 
The results for the scale independent and scale dependent RCs are collected in 
the third column of tables~\ref{table:zeta1} and \ref{table:zeta2} respectively.
The second error, when quoted, represents the systematic uncertainty due to 
residual discretization effects, estimated from the quality of the plateaus of
the RGI combinations.
\renewcommand{\arraystretch}{1.5}
\begin{table}[t]
\begin{center}
\begin{tabular}{||c||c||c||c|c|c||c|c||}
\cline{2-8}
\multicolumn{1}{c||}{{\phantom{\huge{l}}}\raisebox{-.2cm}{\phantom{\Huge{j}}}} 
& & {\sf RI/MOM} & \multicolumn{3}{c||}{WARD IDENTITIES} & 
\multicolumn{2}{c||}{BPT 1-loop} \\
\cline{3-8} 
\multicolumn{1}{c||}{{\phantom{\huge{l}}}\raisebox{-.2cm}{\phantom{\Huge{j}}}} 
& $\beta$ & {\sf This work} & This work & LANL & ALPHA& $c_{SW}=1$ & $c_{SW}^{NP}$ \\
\hline \hline
          &  6.0  & \sf{0.772(2)(2)}& 0.774(4) & 0.770(1) & 0.7809(6)& 0.7820 & 0.8504 \\
$Z_V$     &  6.2  & \sf{0.783(3)}   & 0.789(2) & 0.7874(4)& 0.792(1) & 0.7959 & 0.8463 \\
          &  6.4  & \sf{0.801(2)}   & 0.804(2) & 0.8018(5)& 0.803(1) & 0.8076 & 0.8480 \\
          &  6.45 & \sf{0.803(3)}   &  -----   &  -----   &[0.808(1)]& 0.8103 & 0.8488 \\
\hline \hline
          &  6.0  & \sf{0.812(2)}   & 0.856(17)(15)& 0.807(8) & 0.791(9) & 0.8038 & 0.8693 \\
$Z_A$     &  6.2  & \sf{0.819(3)}   & 0.812(5)(2)  & 0.818(5) & 0.807(8) & 0.8163 & 0.8624 \\
          &  6.4  & \sf{0.832(3)}   & 0.843(10)(1) & 0.827(4) & 0.827(8) & 0.8269 & 0.8628 \\
          &  6.45 & \sf{0.833(3)}   &	-----	   &  -----   &[0.825(8)]& 0.8293 & 0.8633 \\
\hline \hline
          &  6.0  & \sf{0.870(4)(5)}& 0.893(20)(19)& 0.842(5) &[0.840(8)]& 0.9398 & 0.9444 \\
$\dfrac{Z_P}{Z_S}$ 
          &  6.2  & \sf{0.877(5)}   & 0.877(5)(1)  & 0.884(3) &[0.886(9)]& 0.9449 & 0.9545 \\
          &  6.4  & \sf{0.894(3)}   & 0.914(10)(1) & 0.901(5) &[0.908(9)]& 0.9491 & 0.9594 \\
          &  6.45 & \sf{0.897(4)}   &	-----	   &  -----   &[0.912(9)]& 0.9500 & 0.9603 \\
\hline \hline
\end{tabular}
\end{center}
\caption{\label{table:zeta1} \sl \small Values of the scale independent RCs as 
obtained with the $\ri$ method, the WI method and one loop boosted perturbation
theory (BPT).}
\end{table}
\renewcommand{\arraystretch}{1.0}
\renewcommand{\arraystretch}{1.5}
\begin{table}
\begin{center}
\begin{tabular}{||c||c||c||c||c|c||}
\cline{2-6} 
\multicolumn{1}{c||}{{\phantom{\huge{l}}}\raisebox{-.2cm}{\phantom{\Huge{j}}}} 
& & \sf{RI/MOM} & SF &\multicolumn{2}{c||}{BPT 1-loop} \\
\cline{3-6} 
\multicolumn{1}{c||}{{\phantom{\huge{l}}}\raisebox{-.2cm}{\phantom{\Huge{j}}}} 
& $\beta$ & \sf{This work} & ALPHA & $c_{SW}=1$ & $c_{SW}^{NP}$ \\
\hline \hline
                &  6.0  & \sf{0.839(2)(6)}&  -----  & 0.8151 & 0.8821 \\
$Z_q^{RI}$(1/a) &  6.2  & \sf{0.850(2)}   &  -----  & 0.8269 & 0.8751 \\
                &  6.4  & \sf{0.861(2)}   &  -----  & 0.8369 & 0.8750 \\
                &  6.45 & \sf{0.860(3)}   &  -----  & 0.8391 & 0.8754 \\
\hline \hline
                &  6.0  & \sf{0.606(2)(3)}&  -----  & 0.6685 & 0.6257 \\
$Z_S^{RI}$(1/a) &  6.2  & \sf{0.642(3)}   &  -----  & 0.6896 & 0.6558 \\
                &  6.4  & \sf{0.671(4)}   &  -----  & 0.7074 & 0.6794 \\
                &  6.45 & \sf{0.680(7)}   &  -----  & 0.7115 & 0.6846 \\
\hline \hline
                &  6.0  & \sf{0.525(3)(6)}&[0.54(1)]& 0.6248 & 0.5877 \\
$Z_P^{RI}$(1/a) &  6.2  & \sf{0.564(4)}   &[0.57(1)]& 0.6487 & 0.6236 \\
                &  6.4  & \sf{0.600(4)}   &[0.60(1)]& 0.6689 & 0.6498 \\
                &  6.45 & \sf{0.610(8)}   &[0.61(1)]& 0.6735 & 0.6555 \\
\hline \hline
                &  6.0  & \sf{0.881(2)(3)}&  -----  & 0.8417 & 0.9442 \\
$Z_T^{RI}$(1/a) &  6.2  & \sf{0.876(2)}   &  -----  & 0.8518 & 0.9259 \\
                &  6.4  & \sf{0.884(3)}   &  -----  & 0.8603 & 0.9190 \\
                &  6.45 & \sf{0.883(2)}   &  -----  & 0.8622 & 0.9180 \\
\hline \hline
\end{tabular}
\end{center}
\caption{\label{table:zeta2} \sl \small Values of the scale dependent RCs as 
obtained with the $\ri$ method, the SF method and one loop boosted perturbation 
theory (BPT). The results are expressed in the $\ri$ scheme at the scale $\mu=1
/a$.}
\end{table}
\renewcommand{\arraystretch}{1.0}

\subsection{Comparison with results from WIs}
In order to compare the results obtained with the $\ri$ method with those
obtained by using a different non-perturbative approach, we have also computed
the scale independent RCs $Z_V$, $Z_A$ and the ratio $Z_P/Z_S$ by studying the 
lattice chiral WIs~\cite{boch,crisa}. We have performed the calculation at three
of the four values of the lattice coupling considered for the $\ri$ study, 
namely $\beta=6.0$, $6.2$ and $6.4$. The vector and axial vector current 
operators have been improved at $\Oa$ by using the values of the coefficients 
$c_V$ and $c_A$ determined in refs.~\cite{alpha,lanl}.

The RC $Z_V$ of the local vector current has been determined by imposing
\beq
2\rho \sum_{x}\sum_{\vec y}\,\langle P(x)V_0(y)P(0)\rangle = 
\frac{1}{Z_V}\,\sum_{\vec y}\,\langle A_0(y)P(0)\rangle \,,
\label{eq:zv1}
\eeq
where $2\rho=(\sum_{\vec x}\nabla_0\langle A_0(x)P(0)\rangle)/(\sum_{\vec x}
\langle P(x)P(0)\rangle)$. We find that an independent determination based on 
the WI
\beq
2\rho \sum_{x}\sum_{\vec y}\,\langle P(x)V_0(y)A_0(0)\rangle = 
\frac{1}{Z_V}\,\sum_{\vec y}\,\langle A_0(y)A_0(0)\rangle \,, 
\eeq
provides results consistent with those obtained from eq.~(\ref{eq:zv1}) but with
larger statistical errors.

The RC of the axial current $Z_A$ and the ratio $Z_P/Z_S$ have been determined 
by using the flavour non-singlet identities
\beq
2\rho \sum_{x}\sum_{\vec y}\,\langle P(x)V_k(y)A_k(0)\rangle = 
-\frac{Z_V}{Z_A^2}\,\sum_{\vec y}\,\langle V_k(y)V_k(0)\rangle \,+ 
\frac{1}{Z_V}\,\sum_{\vec y}\,\langle A_k(y)A_k(0)\rangle\, ,
\eeq
with $k$ summed over 1,2,3, and 
\beq
2\rho \sum_{x}\sum_{\vec y}\,\langle P(x)S(y)P(0)\rangle = 
\frac{Z_P}{Z_AZ_S}\,\sum_{\vec y}\,\langle P(y)P(0)\rangle \,+ 
\frac{Z_S}{Z_AZ_P}\,\sum_{\vec y}\,\langle S(y)S(0)\rangle
\eeq
respectively.

The results are presented in the fourth column of table~\ref{table:zeta1}. The 
first error quoted in the table is statistical while the second one represents 
the systematic uncertainty introduced by different estimates of the value of the
improvement coefficient $c_A$. At $\beta=6.0$, in particular, the determination 
$c_A=-0.037$ obtained in ref.~\cite{lanl} is in disagreement with $c_A=-0.083$ 
estimated in ref.~\cite{alpha} by approximately a factor two. This difference 
introduces in turn a large uncertainty in the evaluation of $Z_A$ and even more 
of the ratio $Z_P/Z_S$, as can be seen from the systematic errors quoted in 
table~\ref{table:zeta1}. At the larger values of $\beta$ the estimates of $c_A$ 
obtained in refs.~\cite{alpha} and \cite{lanl} are in better agreement and the 
corresponding uncertainty in the evaluation of the RCs with the WI method is 
significantly reduced.

\begin{figure}
\hspace*{-0.6cm}
\begin{tabular}{cc}
\epsfxsize8.0cm\epsffile{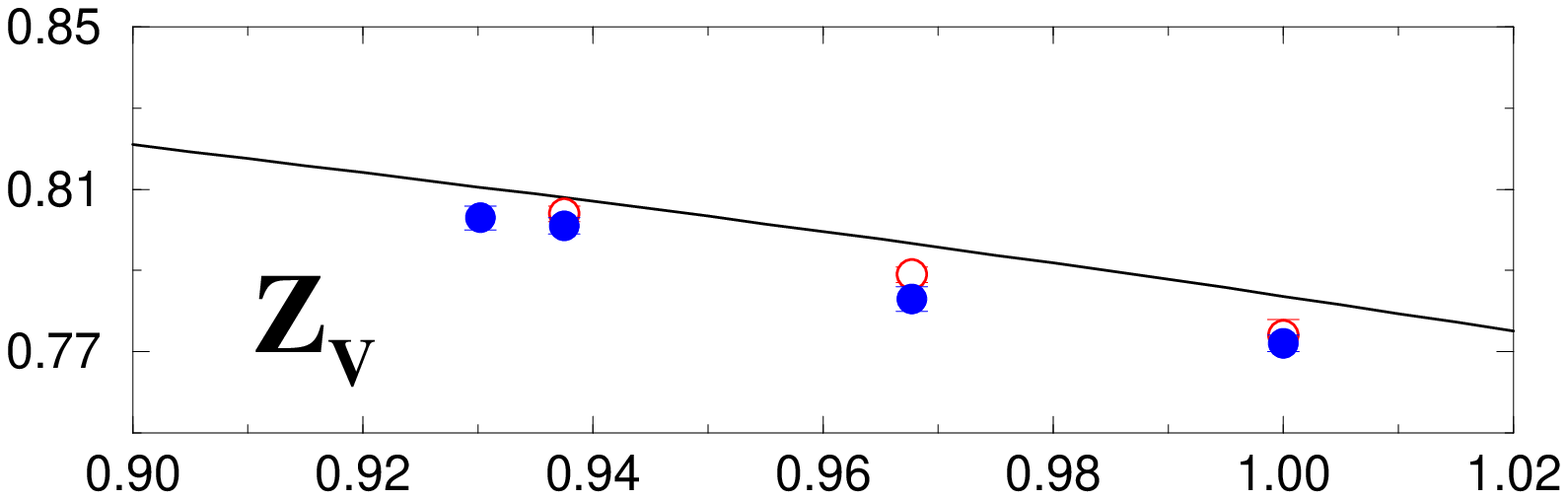} &
\hspace*{-0.5cm}
\epsfxsize8.0cm\epsffile{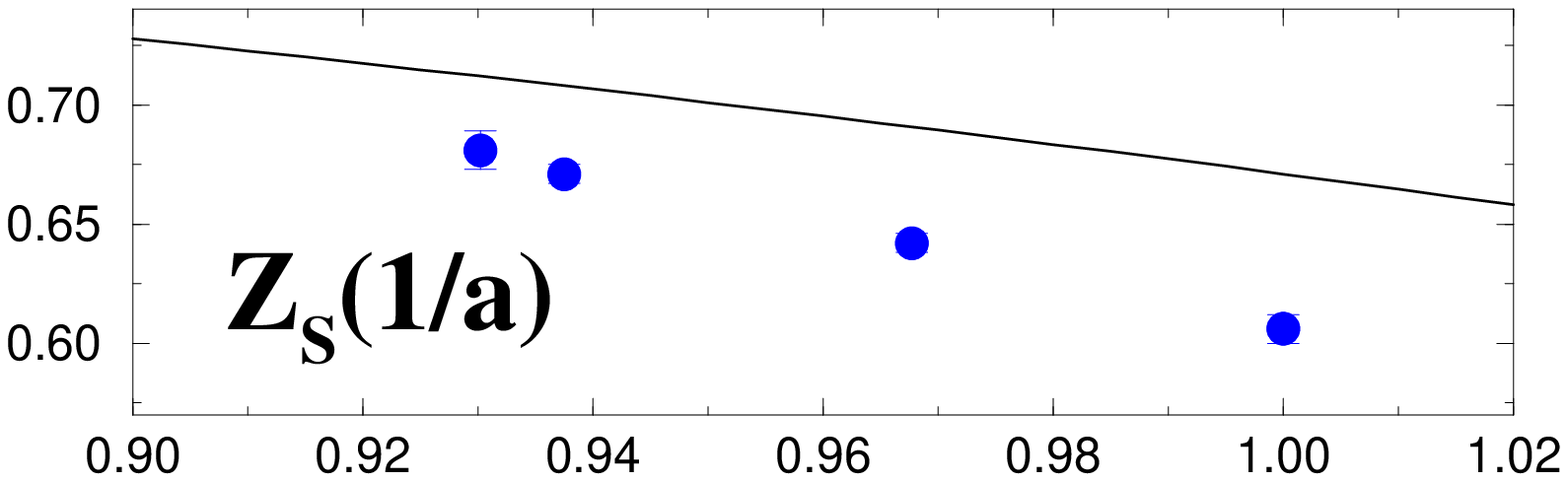} \\
\epsfxsize8.0cm\epsffile{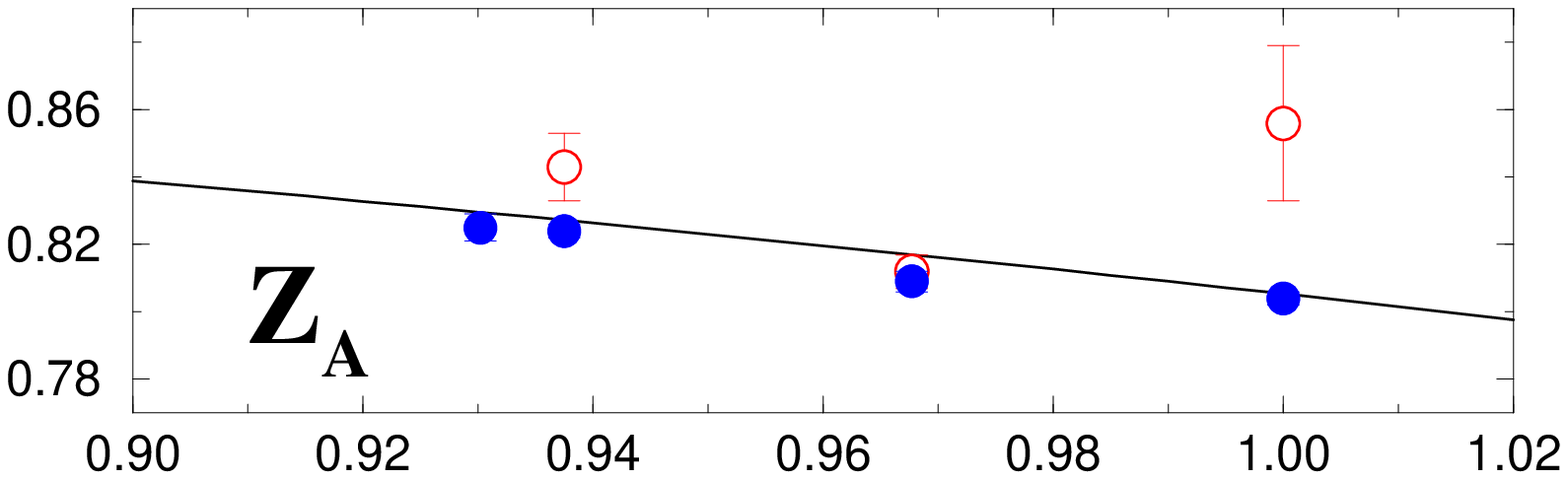} &
\hspace*{-0.5cm}
\epsfxsize8.0cm\epsffile{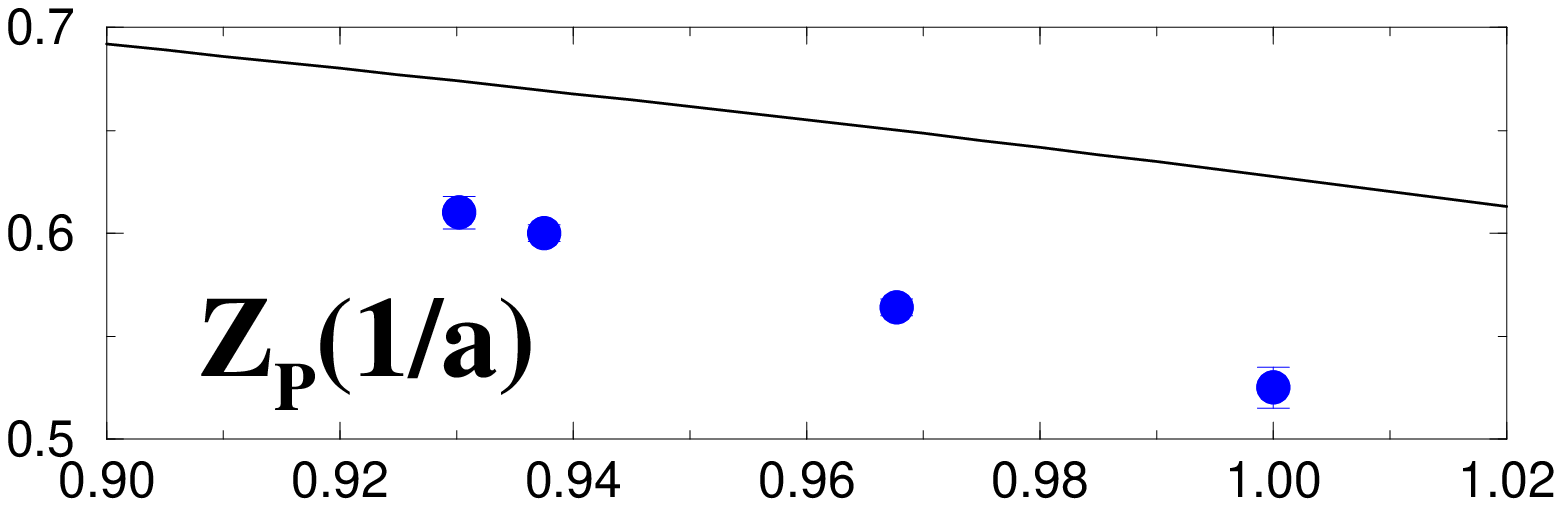} \\
\epsfxsize8.0cm\epsffile{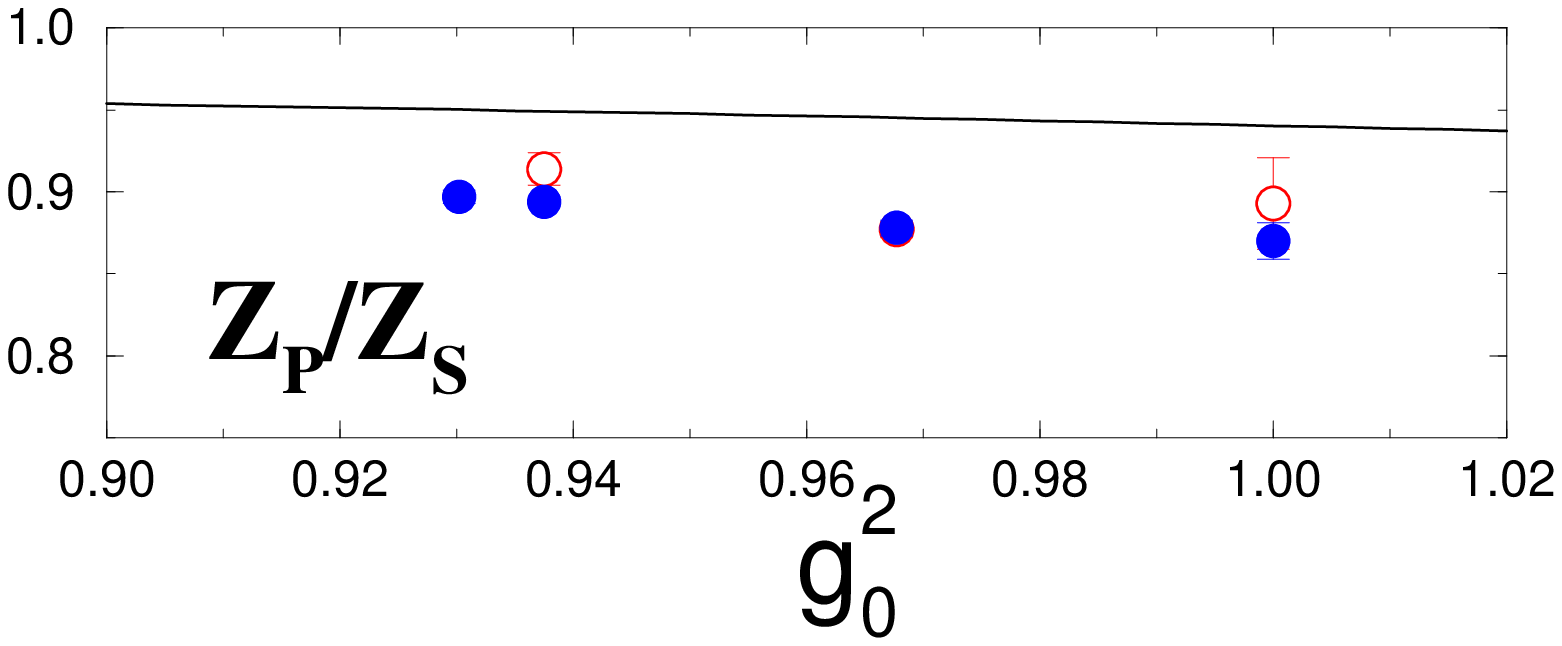} &
\hspace*{-0.5cm}
\epsfxsize8.0cm\epsffile{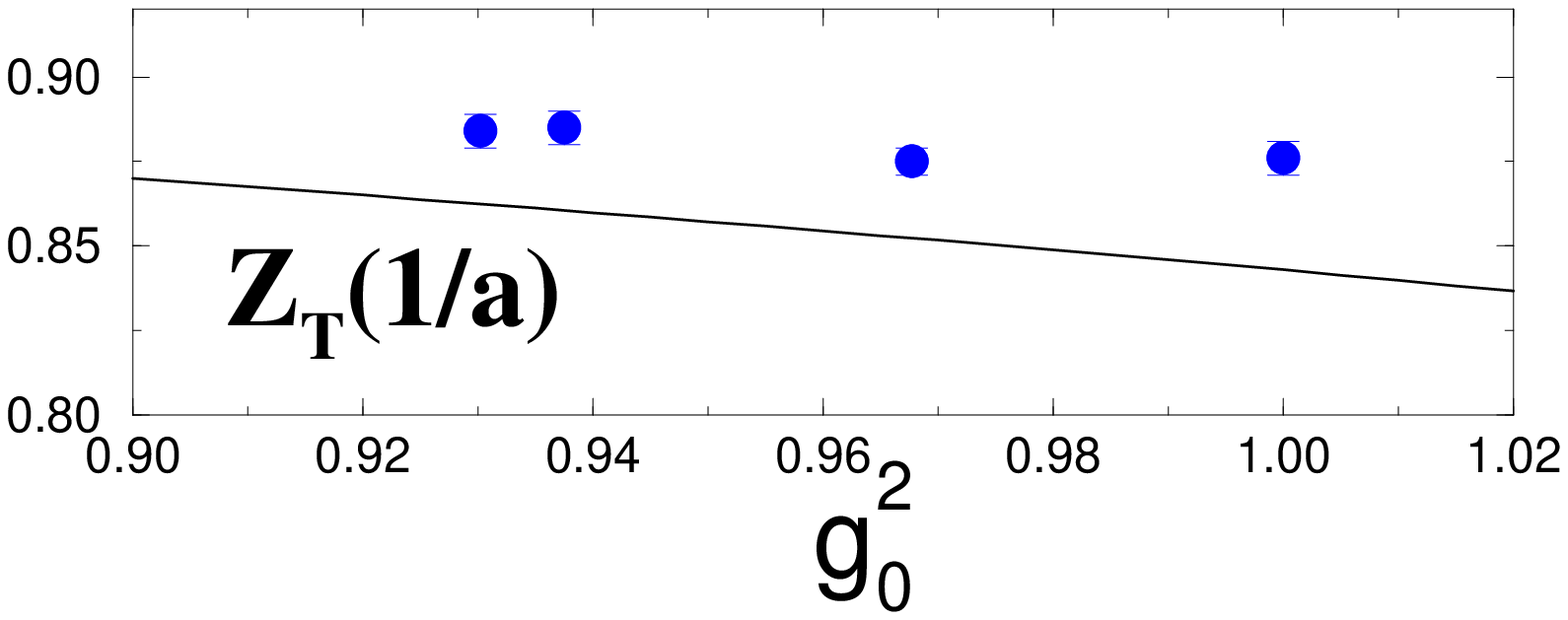} \\
\end{tabular}
\caption{\label{fig:zeta}{\sl \small Values of the RCs as obtained from the 
$\ri$ method (filled circles), the WI method (empty circles) and 1-loop boosted 
perturbation theory with $c_{SW}=1$ (solid lines).}}
\end{figure}

Our final results for the RCs of bilinear quark operators, as obtained from the
$\ri$ and the WI methods (the latter for the scale independent constants) are 
shown in fig.~\ref{fig:zeta}, where they are also compared with the predictions 
of 1-loop boosted perturbation theory.
We observe a very good agreement, typically within one standard deviation, 
between the $\ri$ and the WI determinations. The central values obtained with
the two approaches differ by less than 1\% for $Z_V$, 5\% for $Z_A$ and 3\% for
$Z_P/Z_S$.

\subsection{Comparison with perturbation theory}
The expressions of the RCs of bilinear quark operators computed in perturbation 
theory (PT)~\cite{capitani} are functions of both the coupling constant $g^2$ 
and of $c_{SW}$, the improvement coefficient of the fermionic action. In the 
numerical calculation $c_{SW}$ has been fixed to its non-perturbative 
value~\cite{alpha}. In PT one can use either this non-perturbative estimate or 
the perturbative expression of $c_{SW}$, consistently evaluated at the proper 
order in $g^2$. The difference between the two evaluations represents an 
uncertainty which is of the same order of the other higher order terms neglected
in the perturbative expansion.

In tables~\ref{table:zeta1} and \ref{table:zeta2} we present the predictions for
the RCs obtained by using one-loop boosted PT in two cases, with $c_{SW}$ fixed 
either to its tree-level value\,\footnote{In the perturbative expansion of the 
RCs $c_{SW}$ only enters at ${\cal O}(g^2)$.}, $c_{SW}=1$, or to its 
non-perturbative estimate. The differences between the two sets of results, in
the range $6.0 \le \beta \le 6.45$, are roughly of the order of 10\%. This can
be considered therefore the typical size of uncertainty associated with the
predictions of one-loop boosted PT. The perturbative estimates, shown for
illustration in fig.~\ref{fig:zeta}, are those obtained with $c_{SW}=1$.

The comparison between the non-perturbative $\ri$ determinations and the 
corresponding perturbative estimates at one loop, presented in 
tables~\ref{table:zeta1} and \ref{table:zeta2} and illustrated in 
fig.~\ref{fig:zeta}, shows that the differences are indeed at the expected level
of 10\%. A somewhat larger discrepancy is observed for $Z_P$. In this case, at 
$\beta=6.0$, it ranges between 10\% and 20\%. As expected, we find that in all
cases the agreement between perturbative and non-perturbative determinations of 
the RCs improves by going towards smaller values of the coupling $g^2$.

We conclude this discussion by observing that the differences between 
perturbative and non-perturbative results for the RCs are significantly larger 
than the typical size of statistical errors in present lattice calculations. 
This is the reason why the use of NPR techniques should be considered, at 
present, a fundamental ingredient in the lattice determinations of the physical 
hadronic matrix elements.

\subsection{Comparison with other results in the literature}
The scale independent RCs $Z_V$, $Z_A$ and the ratio $Z_P/Z_S$, for the 
$\Oa$-improved Wilson action considered in this paper, have been also computed 
by the ALPHA~\cite{alpha_zva,alpha_zsp} and LANL~\cite{lanl} Collaborations by 
using the WI method. The results are collected in table~\ref{table:zeta1}. We
have quoted in square brackets the estimates of the RCs which are not given
directly in the original papers but can be inferred from the published results. 
Specifically, the ALPHA results for the RCs at the value of the coupling $\beta
=6.45$ have been obtained by using the fits of these quantities, in terms of 
rational functions of $g^2$, performed in refs.~\cite{alpha_zva,alpha_zsp}. We 
have also quoted in square brackets in table~\ref{table:zeta1} the ALPHA results
for the ratio $Z_P/Z_S$, which have been obtained by combining the results for 
the quantity $Z_P/(Z_A\,Z_S)$ presented in ref.~\cite{alpha_zsp} with the 
estimates of $Z_A$ given in ref.~\cite{alpha_zva}.

The comparison between the $\ri$ results presented in this paper with those
obtained by ALPHA and LANL by using the WI method shows an agreement which is,
in most of the cases, at the level of 1\% or even better. Slightly larger 
differences, at the level of 3\%, are only observed for the estimates of $Z_A$ 
and $Z_P/Z_S$ at $\beta=6.0$.

In table~\ref{table:zeta2} we also present (in square brackets) the results for 
the RC of the pseudoscalar density $Z_P$ obtained by the ALPHA 
Collaboration~\cite{alpha_zm} by using the SF approach. Continuum perturbation
theory at one loop has been used in this case to convert these estimates from 
the SF to the $\ri$ renormalization scheme, at the proper value of the 
renormalization scale. Also in this case, we find an agreement which is at the 
level of 1\% or better. In conclusion, we believe that these comparisons, 
performed among results obtained by different groups and by using different 
approaches, provide strong evidence of the high level of accuracy reached at 
present by the several NPR methods and, in particular, by the $\ri$ approach. 
This also provides us with additional confidence on the quality of the results 
obtained for those RCs, like $Z_S$ and $Z_T$, for which non-perturbative 
estimates have not been presented before with methods different from $\ri$.

\section{Four-fermion operators}
\label{sec:fourf}
We now come to the case of $\Delta F=2$ and $\Delta I=3/2$ four-fermion 
operators which, in a mass independent renormalization scheme, share the same
set of RCs. From a phenomenological point of view these operators play an 
important role since they enter the effective Hamiltonian of weak interactions.
Matrix elements of $\Delta F=2$ operators, for instance, control the $K^0-\bar 
K^0$ and $B^0-\bar B^0$ mixing amplitudes, in both the Standard Model and its 
SUSY extensions (see for instance~\cite{df2susy1,df2susy2}). Operators with 
$\Delta I=3/2$ provide the amplitude of $K\to\pi\pi$ decays in the $I=2$ 
channel, and enter therefore the theoretical estimates of the $\Delta I=1/2$ 
rule and of the direct CP violation parameter $\varepsilon'/\varepsilon$.

Due to the explicit breaking of chiral symmetry induced by the Wilson term,
four-fermion $\Delta F=2$ and $\Delta I=3/2$ operators mix with operators of the
same dimension but of different ``naive'' 
chirality~\cite{boch,marti83}.~\footnote{
In the $SU(2)$ isospin limit there is no mixing with operators of lower 
dimension.} In order to define the complete basis of dimension-six four-fermion 
operators which mix under renormalization we introduce the notation
\beq
O^\pm_{\Gamma \Gamma'} \equiv \frac{1}{2} 
\left [ (\qbar_1\Gamma q_2)(\qbar_3\Gamma' q_4) \pm  
(\qbar_1\Gamma q _4)(\qbar_3\Gamma' q_2) \right ].
\label{eq:qgamgam}
\eeq
and $O_{[\Gamma_1\Gamma_2\pm\Gamma_3\Gamma_4]}\equiv O_{[\Gamma_1\Gamma_2]}\pm
O_{[\Gamma_3\Gamma_4]}$. The basis consists then of ten parity conserving (PC) 
and ten parity violating (PV) operators
\bea
Q^\pm_1  \equiv  O^\pm_{[VV+AA]} && \ct{Q}^\pm_1 \equiv O^\pm_{[VA+AV]}
\nonumber \\
Q^\pm_2  \equiv  O^\pm_{[VV-AA]} && \ct{Q}^\pm_2 \equiv O^\pm_{[VA-AV]}
\nonumber \nonumber \\
\label{eq:qq}	 
Q^\pm_3  \equiv  O^\pm_{[SS-PP]} & \qquad \qquad & 
\ct{Q}^\pm_3  \equiv  - O^\pm_{[SP-PS]}\\
Q^\pm_4  \equiv  O^\pm_{[SS+PP]} && \ct{Q}^\pm_4  \equiv  O^\pm_{[SP+PS]}
\nonumber \\
Q^\pm_5  \equiv  O^\pm_{TT}\qquad && \ct{Q}^\pm_5 \equiv O^\pm_{T \tilde T} 
\nonumber \,.
\eea

As extensively discussed in~\cite{donini}, with Wilson fermions the 
renormalization pattern of the PV sector follows the ``continuum" one
\beq
\left(\begin{array}{c} 
\hat{\ct{Q}}_1 \\ \hat{\ct{Q}}_2 \\ \hat{\ct{Q}}_3 \\ \hat{\ct{Q}}_4 \\ 
\hat{\ct{Q}}_5 \end{array}\right)^\pm =
\left(\begin{array}{rrrrr}
\ct{Z}_{11} & 0 & 0 & 0 & 0 \\
0 & \ct{Z}_{22} & \ct{Z}_{23} & 0 & 0 \\
0 & \ct{Z}_{32} & \ct{Z}_{33} & 0 & 0 \\
0 & 0 & 0 & \ct{Z}_{44} & \ct{Z}_{45} \\
0 & 0 & 0 & \ct{Z}_{54} & \ct{Z}_{55}
\end{array}\right)^\pm
\left(\begin{array}{c} 
\ct{Q}_1 \\ \ct{Q}_2 \\ \ct{Q}_3 \\ \ct{Q}_4 \\ \ct{Q}_5
\end{array}\right)^\pm \, .
\label{eq:renpv}
\eeq
In the PC sector, instead, the explicit breaking of chiral symmetry induces an
additional mixing which is parameterized in terms of a matrix $\Delta$, 
\beq
\left(\begin{array}{c} 
\hat Q_1 \\ \hat Q_2 \\ \hat Q_3 \\ \hat Q_4 \\ \hat Q_5
\end{array}\right)^\pm =
\left(\begin{array}{rrrrr}
Z_{11} & 0 & 0 & 0 & 0 \\
0 & Z_{22} & Z_{23} & 0 & 0 \\
0 & Z_{32} & Z_{33} & 0 & 0 \\
0 & 0 & 0 & Z_{44} & Z_{45} \\
0 & 0 & 0 & Z_{54} & Z_{55}
\end{array}\right)^\pm
\left(\begin{array}{rrrrr}
1 & \Delta_{12} & \Delta_{13} & \Delta_{14} & \Delta_{15} \\
\Delta_{21} & 1 & 0 & \Delta_{24} & \Delta_{25} \\
\Delta_{31} & 0 & 1 & \Delta_{34} & \Delta_{35} \\
\Delta_{41} & \Delta_{42} & \Delta_{43} & 1 & 0 \\
\Delta_{51} & \Delta_{52} & \Delta_{53} & 0 & 1
\end{array}\right)^\pm
\left(\begin{array}{c} 
\tilde Q_1 \\ \tilde Q_2 \\ \tilde Q_3 \\ \tilde Q_4 \\ \tilde Q_5
\end{array}\right)^\pm \, ,
\label{eq:renor_subt}
\eeq
In matrix notation, we can write
\beq
\hat \ct{Q}^\pm = \ct{Z}^\pm \ct{Q}^\pm \qquad , \qquad
\hat Q^\pm = Z^\pm [I+\Delta^\pm]Q^\pm \, ,
\label{eq:PCdecomp}
\eeq
where $I$ is the $5 \times 5$ unit matrix.

\subsection{The RI/MOM method for four-fermion operators}
The procedure implemented to compute non-perturbatively the RCs of four-fermion 
operators within the RI/MOM scheme is a generalization of that explained for 
bilinears. It has been presented in ref.~\cite{donini}. Here we briefly outline 
this procedure and refer the reader to that reference for more details. The 
calculation proceeds in a number of steps:
\begin{enumerate}
\item For each four-fermion operator ${\cal O}_i$, we start by computing the 
four-point Green function in the Landau gauge
\beq\label{eq:unamp}
G_i(x_1,x_2,x_3,x_4)=\langle\,\hat q_1(x_1)\,\hat \qbar_2(x_2)\,{\cal
O}_i(0)\, \hat q_3(x_3)\,\hat \qbar_4(x_4)\,\rangle\ ,
\eeq
where $\hat q$ and $\hat \qbar$ are renormalized quark fields. For convenience,
we take the Fourier transform of $G_i$ with all external quark legs at equal 
momentum $p$. In this way we obtain Green functions in momentum space of the 
form $G_i(p)^{abcd}_{\alpha\beta\gamma\delta}$, where the superscripts and 
subscripts are respectively the colour and spinor indices of the four external 
fields in eq.~(\ref{eq:unamp}).

\item Green functions are then amputated by multiplying $G_i(p)$ by four 
renormalized inverse quark propagators $\hat S^{-1}(p)$:
\beq\label{eq:amp}
\Lambda_i(p)^{a^\prime b^\prime c^\prime d^\prime}_
{\alpha^\prime\beta^\prime\gamma^\prime\delta^\prime}=
\hat S^{-1}(p)^{a^\prime a}_{\alpha^\prime\alpha}\,
\hat S^{-1}(p)^{c^\prime c}_{\gamma^\prime\gamma}\, G_i(p)
^{abcd}_{\alpha\beta\gamma\delta}\,
\hat S^{-1}(p)^{b b^\prime}_{\beta\beta^\prime}\,
\hat S^{-1}(p)^{d d^\prime}_{\delta\delta^\prime}\ .
\eeq
\item Projection operators $P^\pm_i$, with the label $i$ running over all the 
operators of the basis, are introduced which satisfy the orthogonality relation
\beq
\label{eq:projection}
\Tr\{\Lambda_i^{\pm(0)}\,P^\pm_k\}\,=\,\delta_{ik}\ .
\eeq
In the previous equation, the trace is taken over colour and spinor labels and 
$\Lambda_i^{\pm(0)}$ is the tree-level amputated Green function of the operator 
${\cal O}_i$. The explicit form of the projection operators is given in 
ref.~\cite{donini}.

\item Finally, the $\ri$ renormalization conditions are implemented. In the case
of PV operators these conditions take the form
\beq
\left. {\cal Z}^\pm_{ij} \, \Gamma^\pm_{jk}(p)\right|_{p^2=\mu^2} \equiv
\left. {\cal Z}^\pm_{ij} \, \Tr\{\Lambda^\pm_j(p) P^\pm_k\}\right|_{p^2=\mu^2} =
\delta_{ik}\, ,
\label{eq:rengfPV}
\eeq
while for the PC ones they read
\beq
\left. Z^\pm_{il}\,(I+\Delta^\pm)_{lj}\, \Gamma^\pm_{jk}(p)\right|_{p^2=\mu^2}
\equiv \left. Z^\pm_{il}\,(I+\Delta^\pm)_{lj}\, \Tr\{\Lambda^\pm_j(p) P^\pm_k\}
\right|_{p^2=\mu^2} =\delta_{ik}\, .
\label{eq:rengfPC}
\eeq
In order to ensure the mass independence of the renormalization scheme, the 
matrices ${\cal Z}$, $Z$ and $\Delta$ are determined after the extrapolation to 
the chiral limit of the correlation functions entering eqs.~(\ref{eq:rengfPV}) 
and (\ref{eq:rengfPC}).
\end{enumerate}

\subsection{Subtraction of the Goldstone pole}
As discussed in sec.~\ref{sec:syst-goldstone} for the case of the bilinear 
pseudoscalar operator, a possible difficulty in the implementation of the above
procedure comes from the coupling of the operators to the Goldstone boson. In 
the renormalization of the four-fermion operators this difficulty was already 
observed in ref.~\cite{Dawson:2000kh}. 

By using the LSZ reduction formula we show in appendix B that in the PV case 
only a single pion pole can be present, while in the PC case the appearance of a
double pole is also possible. This is the main difference with respect to the
case of the bilinear pseudoscalar operator.

\begin{figure}
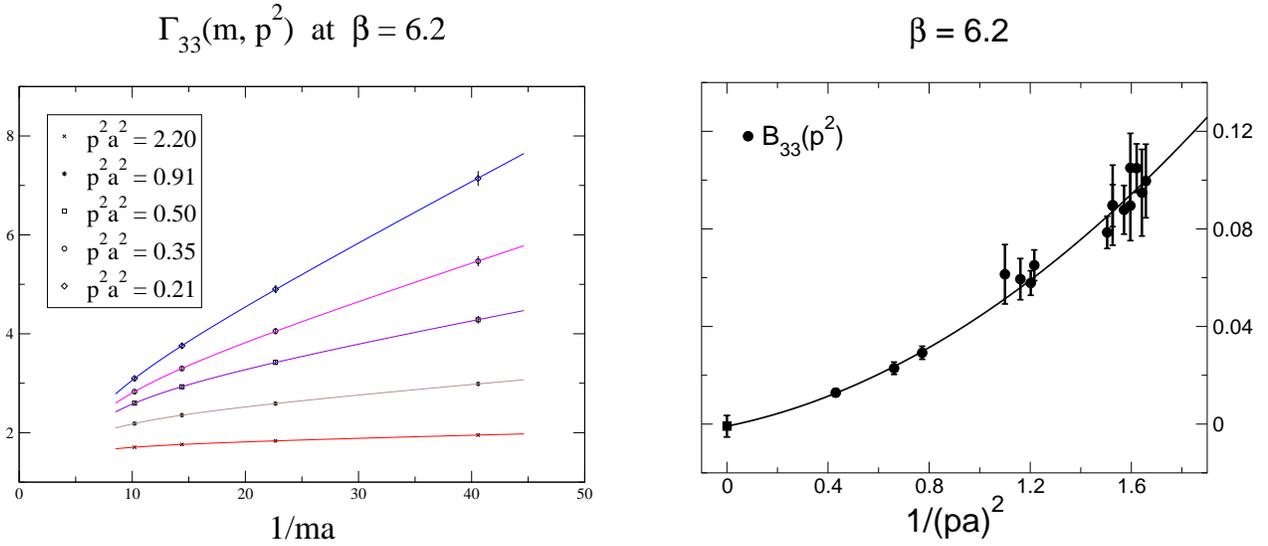

\hspace*{-0.6cm}
\mbox{
\epsfig{figure=inversem.eps,width=0.49\linewidth}
\put(40,0){\epsfig{figure=1sup2.eps,width=0.46\linewidth}}
}
\caption{\sl \small Left: The projected Green function $\Gamma_{33}(m,p^2)$ as a
function of $1/am$ for five different values of $p^2$. The solid lines represent
a fit to eq.~(\ref{eq:fit1}). Right: the coefficient $B_{33}(p^{2})$ of 
eq.~(\ref{eq:fit1}) as a function of $1/(ap)^2$ and the corresponding fit to 
eq.~(\ref{eq:b(p2)}). 
\label{fig:unosuemme1}}
\end{figure}

In the PV case, by performing the OPE we find that the dependence on the quark 
mass $m$ and the external momentum $p$ of the projected Green function $\Gamma$ 
can be written as\footnote{For notational simplicity, we drop from now on, 
unless necessary, the {\it plus} or {\it minus} superscript.} 
\beq
\label{eq:gamij} 
\Gamma_{ij}(m, p^{2}) = \bar  \Gamma_{ij}(m, p^{2}) +
\frac{\Delta \Gamma_{ij}(m,p^2)} {m\, p^{2}} + \dots \, , 
\eeq 
where dots represent non-perturbative contributions which vanish at large 
momenta as $1/p^{4}$ or faster. In eq.~(\ref{eq:gamij}), $\bar\Gamma$ is the 
short-distance form factor, from which we want to extract the matrix ${\cal Z}$ 
of RCs, while the term proportional to $\Delta\Gamma$ is the contribution from 
the Goldstone boson propagator. An example of this behaviour is shown in the 
left plot of fig.~\ref{fig:unosuemme1}, where the Green function $\Gamma_{33}
(m,p^2)$ (the operator ${\cal Q}_3$ happens to be strongly coupled to the 
Goldstone boson) is plotted as a function of $1/m$ for different values of 
$p^2$. A linear dependence of $\Gamma_{33}$ on $1/m$ is clearly visible in the 
plot, with a slope which is a decreasing function of the external momentum.

In order to subtract the Goldstone boson contribution and to compute the RCs 
directly from the value of $\bar\Gamma$ in the chiral limit we follow the same 
strategy discussed in sec.~\ref{sec:syst-goldstone} for the bilinear 
pseudoscalar operator. This strategy is based on two different approaches. 

In the first case we study the dependence on the quark mass of the projected 
Green functions $\Gamma_{ij}$ by performing a fit to the form 
\beq
\Gamma_{ij}(m, p^{2}) = A_{ij}(p^{2}) +\frac{ B_{ij}(p^{2})}{m}+ C_{ij}(p^{2}) 
\, m\, .
\label{eq:fit1} 
\eeq 
The coefficient $A(p^{2})$ gives a determination of $\bar\Gamma$ in the chiral
limit, whereas $B(p^{2})$ is associated with the Goldstone boson contribution
proportional to $\Delta \Gamma$. The term proportional to $C(p^{2})$ comes from 
the expansion of $\bar\Gamma$ and $\Delta\Gamma$ in the quark mass. The results
of the fit to eq.~(\ref{eq:fit1}), in the case of the correlation function 
$\Gamma_{33}$, are shown as solid lines in the left plot of 
fig.~\ref{fig:unosuemme1}.
The matrix ${\cal Z}$ of RCs is then extracted from the coefficient $A(p^{2})$
at large $p^{2}$.

As an additional check of this analysis, we study the dependence of the 
coefficient $B(p^{2})$ on the external momentum. In the right panel of 
fig.~\ref{fig:unosuemme1} we plot $B_{33}(p^{2})$ at $\beta=6.2$ as a function 
of $1/(ap)^{2}$ for $(ap)^2 \geq 0.6$. The solid line in the figure represents 
the result of the fit to the form
\beq 
B_{33}(p^{2}) = \alpha +\frac{\beta}{(ap)^{2}}+\frac{\gamma}{(ap)^{4}}\, .  
\label{eq:b(p2)} 
\eeq
We find that, within the errors, eq.~(\ref{eq:b(p2)}) gives a good description 
of the data with $\alpha=-0.0009(44)$, a result which is well compatible with 
zero. This is consistent with the expectation that the Goldstone pole can only 
appear in power suppressed non-perturbative contributions.

The second approach used to subtract the Goldstone pole is the one suggested in 
ref.~\cite{gv} and also discussed in sec.~\ref{sec:syst-goldstone} for
the bilinear pseudoscalar operator. Applied to the case of the four-fermion 
operators, this method consists in eliminating the contribution of the Goldstone
boson by constructing the combinations
\bea 
\Gamma_{ij}^{\rm SUB}(m_{1}, m_{2}, p^{2}) =\frac{ m_{1} \,\Gamma_{ij}(m_{1}, 
p^{2}) -m_{2} \, \Gamma_{ij}(m_{2}, p^{2})}{m_{1} -m_{2} }  \, . 
\label{eq:gvratio}
\eea 
The subtracted correlation functions are then extrapolated to the chiral limit
by performing a fit to the form 
\beq 
\Gamma_{ij}^{\rm SUB}(m_{1}, m_{2},p^{2}) = \tilde A_{ij}(p^{2}) + 
\tilde C_{ij}(p^{2}) (m_{1} + m_{2}) \, . 
\label{eq:gvfit} 
\eeq 
We find that the results for $\tilde A(p^{2})$, obtained from a fit which uses 
only the four lightest values of quark masses, differ from those of $A(p^{2})$ 
of eq.~(\ref{eq:fit1}) by less than 1\% in all the range of $p^{2}$ considered 
in this study. Therefore, at this level of accuracy, the two approaches provide
the same estimate of the RCs.

In the PC case also a double Goldstone boson pole can be present. For this
reason, we fit in this case the subtracted correlation function of 
eq.~(\ref{eq:gvratio}) to the form 
\beq 
\Gamma_{ij}^{\rm SUB}(m_{1}, m_{2},p^{2}) = \tilde A_{ij}(p^{2}) + 
\tilde C_{ij}(p^{2})(m_{1}+m_{2})+\frac{\tilde E_{ij}(p^{2})}{m_{1} m_{2}}\, ,
\label{eq:gvfitPC}
\eeq
which differs from eq.~(\ref{eq:gvfit}) for the presence of the last term. As 
before, we compute the RCs from the coefficient $\tilde A(p^{2})$. Had we used 
this form in the PV case, a value of $\tilde E(p^{2})$ fully compatible with 
zero would have been obtained. In the PC case instead, our numerical results 
clearly show the presence of such contribution, which is however much smaller 
than the single pole contribution proportional to $B(p^2)$. 

\begin{figure}
\begin{center}
\epsfig{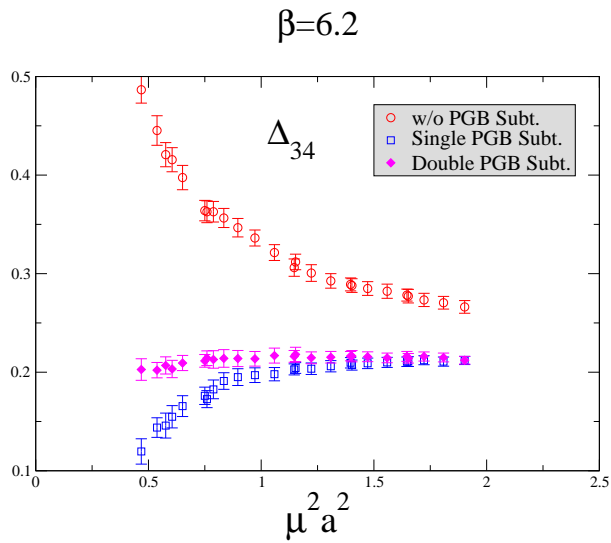}
\caption{\sl \small $\Delta_{34}$ as function of $\mu$ at $\beta=6.2$ in three 
cases: without the subtraction of the Goldstone pole (circles), with the 
subtraction of the single pole (squares) and with the subtraction of both the 
single and the double poles (diamonds). 
\label{fig:deltas}}
\end{center}
\end{figure}
The effect of single and double poles, when not subtracted, is clearly visible 
in several matrix elements, both of the scale dependent RCs, ${\cal Z}(\mu a,
g_0^2)$ and $Z(\mu a,g_0^2)$, and of the scale independent mixing coefficients
$\Delta(g_0^2)$. An example of the latter case, namely the one of the matrix 
element $\Delta_{34}$, is shown in fig.~\ref{fig:deltas}. It is very reassuring 
that $\Delta_{34}$ becomes almost independent of the scale, as should be the 
case, once the pole contribution has been eliminated. For some other matrix
elements, and in particular in the case of $Z_{11}$ and of the corresponding 
mixing coefficients $\Delta_{1k}$ which are relevant for the lattice estimates 
of $K^0-\bar K^0$ and $B^0-\bar B^0$ mixing in the Standard Model, the effect of
the subtraction of the Goldstone pole is absolutely negligible.

\subsection{Renormalization scale dependence, discretization and finite volume 
effects}
In order to investigate the presence of discretization effects and to study the
renormalization scale dependence of the RCs of four-fermion operators we follow
a procedure close to the one described in sects.~\ref{sec:syst-oa} and 
\ref{sec:syst-allbeta} for the case of bilinear quark operators.

We start by constructing the RGI combinations 
\beq 
\label{eq:appdRGI} 
Z^{RGI} = C^{-1}(\mu){Z}(\mu) 
\eeq 
for both the PC and PV operators, where the evolution matrix $C(\mu)$, for the 
complete basis has been computed, in the $\ri$ scheme, at the NLO in 
perturbation theory~\cite{Ciuchini:1997bw}.

We then rescale the RGI combinations, computed at the different values of the
lattice spacing, to a common reference scale $\bar a$, which in this case we 
choose to be the value of the lattice spacing at $\beta=6.45$. To that purpose,
we compute the renormalization scale independent ratios
\beq
R(a,\bar a)=Z^{RGI}(a)^{-1} Z^{RGI}(\bar a) = Z(a,\mu)^{-1} Z(\bar a,\mu)\, .
\eeq
which are the analogous of those defined in eq.~(\ref{eq:erre}) for bilinear
operators. We then use these ratios to rescale the RGI RCs at the three values 
of the coupling different from $\beta=6.45$.

The rescaled RGI combinations should be independent of both the renormalization
scale and the lattice spacing. This is true, however, only up to discretization
effects and higher-order perturbative corrections not taken into account in the 
NLO perturbative evaluation of the function $C(\mu)$. A comparison of the 
rescaled RGI combinations obtained at the same value of the renormalization 
scale, but at different values of the lattice spacing, provides an estimates of 
discretization effects. On the other hand, by studying at fixed lattice spacing
the dependence of the RGI combinations on the renormalization scale we can 
investigate the effect of higher orders perturbative contributions not included 
in the evaluation of the evolution matrix $C(\mu)$.

\begin{figure}
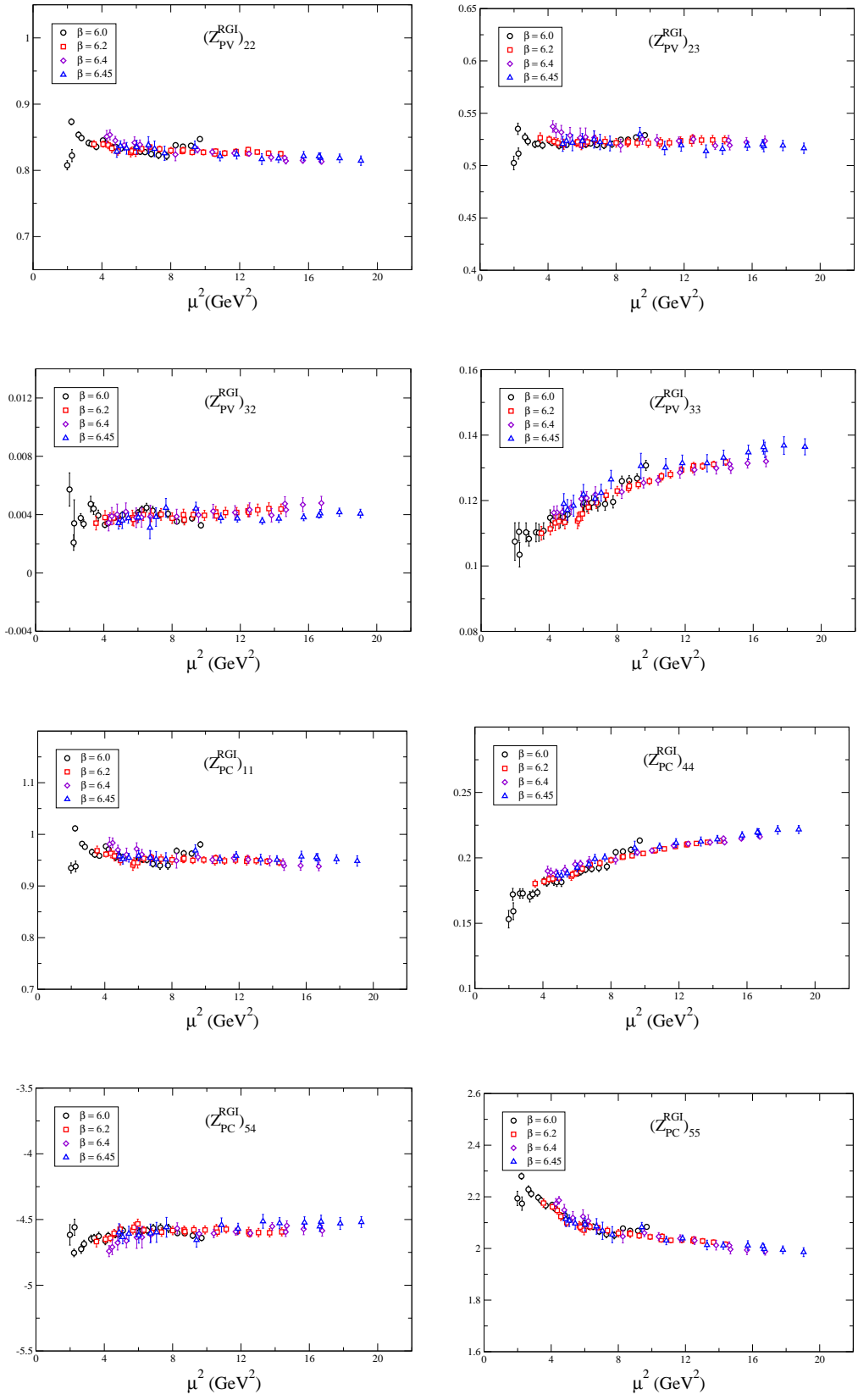

\begin{center}
\vspace*{-1.2cm}\hspace*{-7cm}\mbox{
\epsfig{figure=ratio2_RGI_q2.eps,width=0.39\linewidth}
\put(20,0){\epsfig{figure=ratio3_RGI_q2.eps,width=0.39\linewidth}}}\\
\vspace*{0.78cm}\hspace*{-7.2cm}
\mbox{\epsfig{figure=ratio4_RGI_q2.eps,width=0.40\linewidth}
\put(21,0){\epsfig{figure=ratio5_RGI_q2.eps,width=0.39\linewidth}}}\\
\vspace*{0.78cm}\hspace*{-7cm}
\mbox{\epsfig{figure=ratio1_RGI_PC_q2.eps,width=0.38\linewidth}
\put(20,0){\epsfig{figure=ratio6_RGI_PC_q2.eps,width=0.39\linewidth}}}\\
\vspace*{0.78cm}\hspace*{-7cm}
\mbox{\epsfig{figure=ratio8_RGI_PC_q2.eps,width=0.39\linewidth}
\put(24,0){\epsfig{figure=ratio9_RGI_PC_q2.eps,width=0.38\linewidth}}}
\caption{\sl \small Results for some of the RGI combinations as obtained at the 
four values of the lattice coupling and rescaled to $\beta=6.45$.}
\label{fig:RGI} 
\end{center}
\end{figure}
In fig.~\ref{fig:RGI} we show the numerical results for the rescaled RGI 
combinations, in both the PV and PC cases, as a function of the renormalization 
scale. In these plots we only show the results for the {\it plus} sector. For 
the {\it minus} sector the situation is similar. As can be seen from the plots, 
most of the matrix elements of both $Z^{RGI}$ and ${\cal Z}^{RGI}$ do not show 
any significant dependence on the renormalization scale. There are three cases, 
however, namely the matrix elements 33, 44 and 55, in which the plateau is worse
than in the others. Since deviations from the expected constant behaviour are 
observed in a similar way at all values of the lattice spacing, and the results 
obtained at different $\beta$ are in good agreement among each others, we 
conclude that lattice artifacts are quite small and interpret these deviations 
as mainly due to the effect of N$^2$LO perturbative corrections not included in 
the evaluation of the function $C(\mu)$. This explanation is also supported by 
the fact that the operators $Q_3$, $Q_4$, $Q_5$ and ${\cal Q}_3$, ${\cal Q}_4$, 
${\cal Q}_5$ have very large anomalous dimensions at the leading order. We
emphasize that a better control of the renormalization scale dependence of these
operators could be achieved either by a perturbative calculation of the N$^2$LO 
anomalous dimensions or, even better, by a non-perturbative study of the running
by using an iterative matching technique which involves several lattice scales.
This approach, which has been already implemented within the context of the SF
scheme (see for instance ref.~\cite{alpha_zm} for a study of the pseudoscalar RC
$Z_P$) can be implemented with other NPR techniques as well, like the $\ri$ or
the $x$-space method.

\begin{figure}
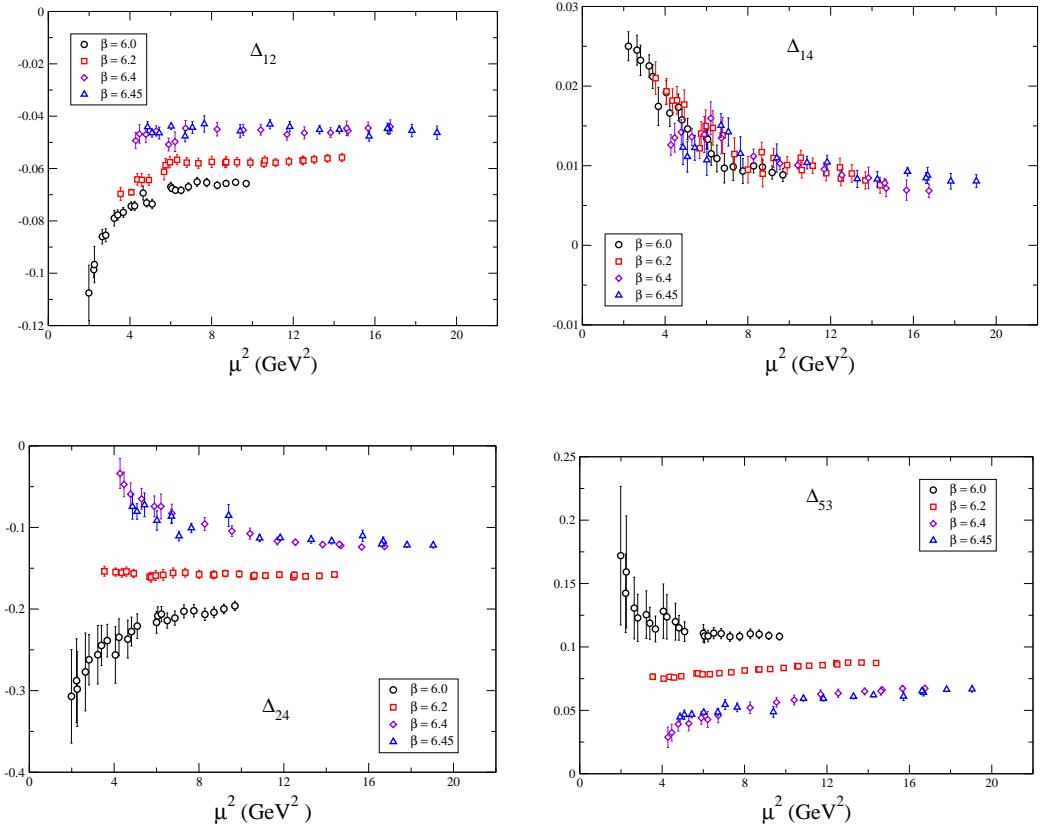

\begin{center}
\vspace*{-1.cm}\hspace*{-7cm}
\mbox{\epsfig{figure=delta01_q2.eps,width=0.4\linewidth}
\put(20,0){\epsfig{figure=delta03_q2.eps,width=0.405\linewidth}}}\\
\vspace*{0.8cm}\hspace*{-7.2cm}
\mbox{\epsfig{figure=delta06_q2.eps,width=0.41\linewidth}
\put(21,0){\epsfig{figure=delta16_q2.eps,width=0.40\linewidth}}}
\caption{\sl \small Results for some of the mixing coefficients $\Delta_{ij}$ 
at the four values of lattice coupling.}
\label{fig:DELTA} 
\end{center}
\end{figure}

In fig.~\ref{fig:DELTA} we show the results for the mixing coefficients $\Delta$
as a function of the renormalization scale $\mu$. At fixed lattice spacing, 
these quantities are expected to be independent of $\mu$. Indeed, in most of 
the cases we observe from the plots a reasonably good scale independence at 
large $p^2$, in particular for those coefficients of relatively large magnitude.
The systematic error induced by the effect of a residual scale dependence is 
accounted for in the final determination of the RCs.
\begin{figure}
\begin{center}
\vspace*{-1.cm}\hspace*{-7cm}
\mbox{
\epsfig{figure=DELTA_cont_lim.eps,width=0.45\linewidth}
\put(20,0){\epsfig{figure=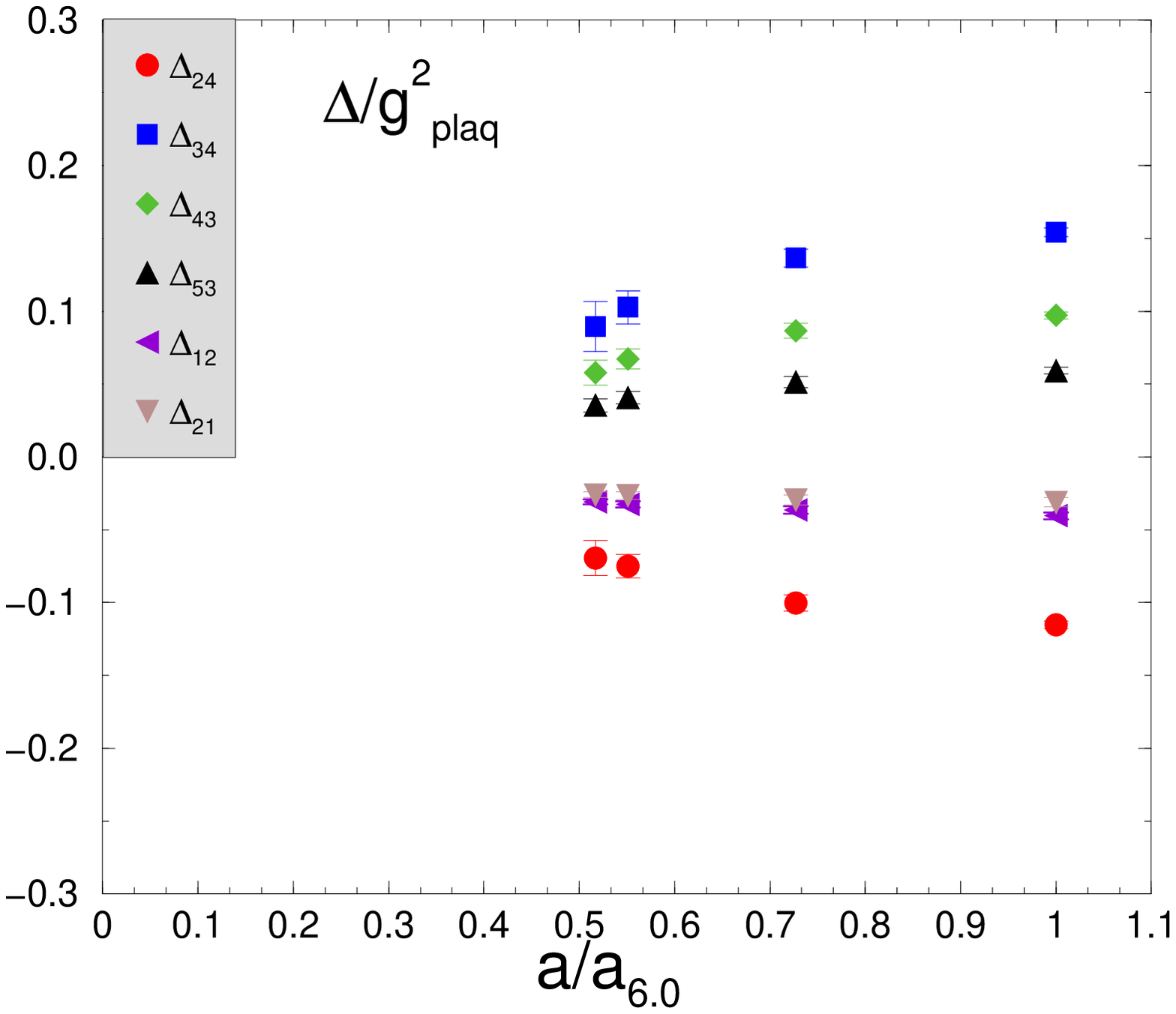,width=0.40\linewidth}}}
\caption{\sl \small Results for some of the matrix elements of $\Delta$ (left) 
and for the ratios $\Delta/g^2_{plaq}$ (right), where $g^2_{plaq}=g^2/(\frac{1}
{3}{\rm Tr}(U_P))$ is the boosted coupling defined from the plaquette, as a
function of the lattice spacing.
\label{fig:deltax}}
\end{center}
\end{figure}

Mixing among operators of different naive chirality, parameterized by the matrix
$\Delta$, is a consequence of the explicit chiral symmetry breaking induced by 
the Wilson term. Therefore this mixing is expected to disappear in the 
continuum limit. To verify this expectation, we show in the left plot of
fig.~\ref{fig:deltax} the results for some of the matrix elements of $\Delta$ as
a function of the lattice spacing. We see from this plot that the absolute 
values of the $\Delta$s decrease as the continuum limit is approached.
In the right panel of fig.~\ref{fig:deltax} we present the same matrix elements 
of $\Delta$ divided by $g^2_{plaq}$, where $g^2_{plaq}=g^2/(\frac{1}{3}{\rm Tr}
(U_P))$ is the boosted coupling defined from the plaquette. If the perturbative 
expansion in terms of $g^2_{plaq}$ is rapidly convergent, the ratios $\Delta/
g^2_{plaq}$ are expected to be flatter than the $\Delta$s themselves. The plot 
in fig.~\ref{fig:deltax} shows that this is indeed the case. The residual 
dependence on the lattice spacing, observed in the figure, signals the presence 
either of higher orders in the perturbative expansion or of finite lattice 
spacing effects. 

\begin{figure}[t]
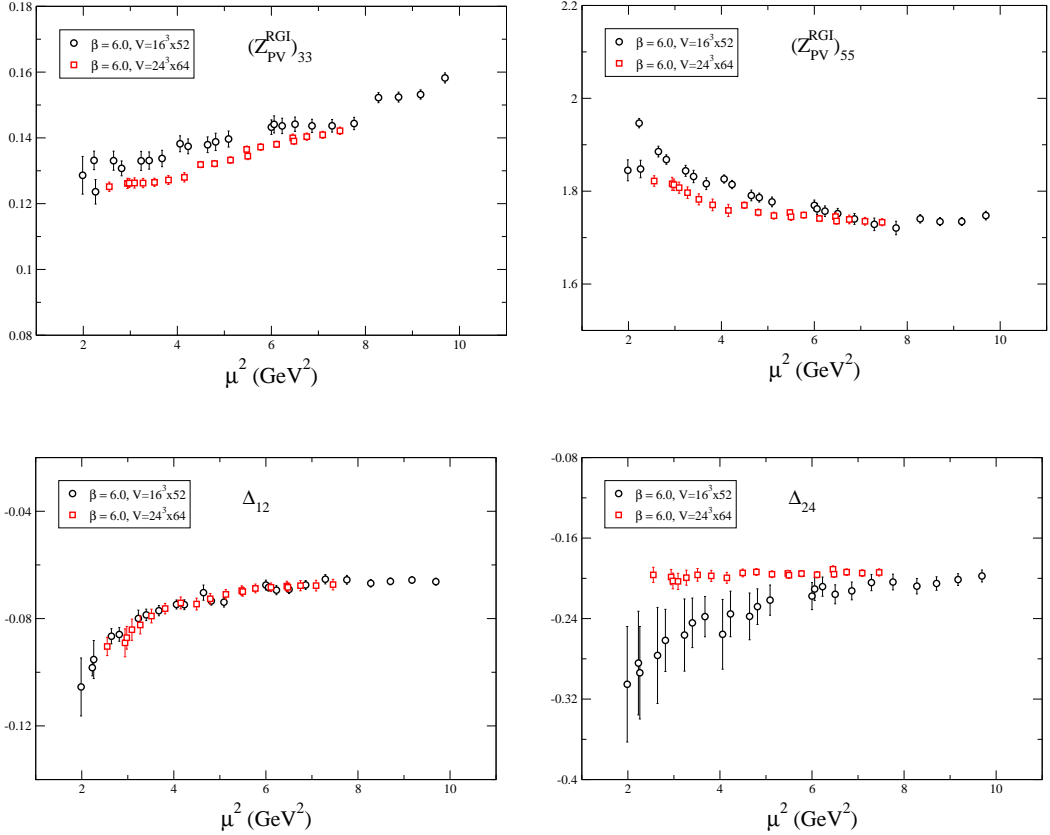

\begin{center}
\vspace*{-1.cm}\hspace*{-7cm}
\mbox{\epsfig{figure=ratio5_RGI_FV_q2.eps,angle=0,width=0.415\linewidth}
\put(20,2){\epsfig{figure=ratio9_RGI_FV_q2.eps,angle=0,width=0.405\linewidth}}}\\
\vspace*{0.8cm}\hspace*{-7.2cm}
\mbox{\epsfig{figure=delta01_FV_q2.eps,angle=0,width=0.41\linewidth}
\put(20,0){\epsfig{figure=delta06_FV_q2.eps,angle=0,width=0.41\linewidth}}}
\caption{\sl \small Comparison for some of the matrix elements of the RCs and 
mixing coefficients as obtained on two different volumes at $\beta=6.0$.}
\label{fig:z4f_FV} 
\end{center}
\end{figure}

To conclude the analysis of the systematic errors which may affect the 
determination of the RCs of four-fermion operators we investigate the presence 
of finite volume effects. As already done for bilinear quark operators (see 
sec.~\ref{sec:syst-vol}), we compare the results for the RCs obtained at $\beta
=6.0$ on two different volumes, namely $16^3\times 52$ and $24^3\times 64$. The
parameters of the simulation on the smallest volume are those given in 
table~\ref{tab:details}. For the calculation on the largest volume, we have used 
a set of 480 gauge field configurations and computed quark propagators at four
values of the Wilson parameter, namely $\kappa =$0.13180, 0.13280, 0.13770, 
0.13440 ($\kappa_{cr}=0.135220(6)$). The comparison of the results is 
illustrated in fig.~\ref{fig:z4f_FV} for some of the matrix elements of the RCs 
and mixing coefficients.
We see from the plots that, in some cases, the presence of finite volume effects
is visible, though they become negligible at large values of the renormalization
scale ($\mu\simge 2.5\gev$). Since our final determination of the RCs is 
obtained from a fit in the interval $\mu=[2.5,3.0]\gev$ (see below), we find
that the results obtained on the different volumes are compatible within the
statistical and the estimated systematic errors. For this reason, we do not 
include in the final error any additional uncertainty due to final volume 
effects, although this point would require more investigations. In order to
better illustrate this comparison, when quoting our final results (see 
tables~\ref{tab:PC+}-\ref{tab:PV-}) we will present at $\beta=6.0$ the
determinations obtained on both the lattices with different size.

\subsection{Results}
In order to obtain the RCs at a given scale we proceed in the following way. We 
extract the RGI combinations $Z^{RGI}$ and ${\cal Z}^{RGI}$ by fitting to a 
constant the plateau in fig.~\ref{fig:RGI}, and similarly for the other matrix 
elements, in the interval $\mu=[2.5,3.0] \gev$. We then use the renormalization 
group evolution at the NLO to obtain the value of the RCs at the desired scale.
A standard choice of the scale, at which matrix elements are matched with Wilson
coefficients in the effective weak Hamiltonian, is $\mu=2\gev$. Since this value
lies in most of the cases within the range of momenta at which the RCs are 
directly computed, we estimate the systematic error in the following way: we 
take the RCs computed non-perturbatively at a given scale $\mu_0$ and run 
them up or down to the desired scale $\mu=2\gev$. The systematic error is then 
estimated from the deviations of the results obtained by performing different 
choices of $\mu_0$. The same procedure has been also applied for the 
determinations of the mixing coefficients $\Delta$s. The only difference is 
that, being free of anomalous dimension, these quantities do not evolve with the
renormalization scale.

We emphasize that the systematic error induced by the renormalization group 
running of the RCs should be regarded as an uncertainty related mainly to the 
perturbative estimates of the Wilson coefficients of the effective Hamiltonian,
rather than to the non-perturbative the determination of the RCs. Indeed, had we
chosen to quote the values of the RCs at the renormalization scale at which they
are directly determined in the lattice calculation, this error would be absent.

Our final estimates of the RCs of four-fermion operators in the $\ri$ scheme at 
the scale $\mu=2\gev$ are collected in tables~\ref{tab:PC+}, \ref{tab:PC-},
\ref{tab:PV+} and \ref{tab:PV-} for the PC$^\pm$ and PV$^\pm$ respectively. As
mentioned before, for the case of $\beta=6.0$, we present the results obtained 
on lattices with two different sizes.
\begin{table}
\begin{center}
\begin{tabular}{|c|c|c|c|c|c|}
\hline
$\beta$&$6.0\,(16^3\times52)$&$6.0\,(24^3\times64)$&$6.2$&$6.4$ & $6.45$ \\
\hline
\hline

\rule[-3mm]{0mm}{8mm}$Z_{11}$&$ 0.609( 4)(13)$&$ 0.604( 3)( 4)$&$ 0.634( 5)( 5)$&$ 0.671(10)( 9)$&$ 0.682( 7)( 5)$ \\
\hline
$\Delta_{12}$&-0.067( 1)( 4)&-0.068( 2)( 4)&-0.058( 2)( 4)&-0.048( 1)( 3)&-0.045( 2)( 2)\\
$\Delta_{13}$&-0.021( 1)( 1)&-0.022( 1)( 1)&-0.016( 1)( 2)&-0.012( 1)( 2)&-0.012( 1)( 2)\\
$\Delta_{14}$& 0.010( 1)( 3)& 0.014( 1)( 3)& 0.011( 2)( 4)& 0.013( 1)( 3)& 0.013( 1)( 3)\\
$\Delta_{15}$& 0.004( 1)( 3)& 0.003( 1)( 2)& 0.006( 1)( 1)& 0.004( 2)( 1)& 0.006( 1)( 3)\\
\hline
\hline

\rule[-3mm]{0mm}{8mm}$Z_{22}$&$ 0.669( 3)(11)$&$ 0.662( 3)( 3)$&$ 0.691( 3)( 5)$&$ 0.732( 8)(10)$&$ 0.740( 8)( 7)$ \\

\rule[-3mm]{0mm}{8mm}$Z_{23}$&$ 0.274( 2)( 8)$&$ 0.276( 2)( 2)$&$ 0.298( 4)( 7)$&$ 0.323( 8)( 9)$&$ 0.328( 6)(10)$ \\
\hline
$\Delta_{21}$&-0.050( 1)( 6)&-0.053( 1)( 5)&-0.047( 2)( 5)&-0.040( 1)( 4)&-0.038( 1)( 3)\\
$\Delta_{24}$&-0.199( 4)( 7)&-0.194( 3)( 3)&-0.158( 6)( 7)&-0.111( 2)(12)&-0.102( 4)(17)\\
$\Delta_{25}$& 0.020( 1)( 2)& 0.019( 1)( 1)& 0.013( 1)( 3)& 0.008( 1)( 3)& 0.008( 1)( 4)\\
\hline
\hline

\rule[-3mm]{0mm}{8mm}$Z_{32}$&$ 0.029( 1)( 1)$&$ 0.028( 1)( 1)$&$ 0.025( 1)( 1)$&$ 0.024( 2)( 1)$&$ 0.022( 1)( 1)$ \\

\rule[-3mm]{0mm}{8mm}$Z_{33}$&$ 0.390( 3)(31)$&$ 0.376( 2)(15)$&$ 0.357( 4)(19)$&$ 0.344( 5)(12)$&$ 0.331( 6)(18)$ \\
\hline
$\Delta_{31}$& 0.014( 1)( 4)& 0.015( 1)( 3)& 0.015( 1)( 3)& 0.013( 1)( 1)& 0.013( 0)( 3)\\
$\Delta_{34}$& 0.269( 6)(18)& 0.260( 4)( 3)& 0.215( 7)( 7)& 0.153( 3)(16)& 0.131( 7)(24)\\
$\Delta_{35}$&-0.009( 0)( 1)&-0.008( 1)( 1)&-0.006( 1)( 1)&-0.003( 1)( 2)&-0.002( 1)( 5)\\
\hline
\hline

\rule[-3mm]{0mm}{8mm}$Z_{44}$&$ 0.478( 3)(33)$&$ 0.463( 2)(15)$&$ 0.445( 4)(20)$&$ 0.434( 5)(14)$&$ 0.419( 5)(21)$ \\

\rule[-3mm]{0mm}{8mm}$Z_{45}$&$-0.024( 1)( 1)$&$-0.023( 1)( 1)$&$-0.019( 1)( 1)$&$-0.017( 2)( 2)$&$-0.015( 2)( 1)$ \\
\hline
$\Delta_{41}$& 0.005( 0)( 2)& 0.005( 0)( 2)& 0.006( 0)( 2)& 0.005( 0)( 1)& 0.005( 0)( 1)\\
$\Delta_{42}$& 0.007( 0)( 2)& 0.006( 0)( 0)& 0.006( 1)( 0)& 0.005( 1)( 2)& 0.003( 1)( 3)\\
$\Delta_{43}$& 0.174( 4)( 6)& 0.164( 2)( 3)& 0.136( 4)( 7)& 0.100( 2)(10)& 0.085( 5)(11)\\
\hline
\hline

\rule[-3mm]{0mm}{8mm}$Z_{54}$&$-0.253( 2)(13)$&$-0.257( 2)( 6)$&$-0.291( 3)(10)$&$-0.329( 6)(13)$&$-0.337( 6)(13)$ \\

\rule[-3mm]{0mm}{8mm}$Z_{55}$&$ 0.786( 4)(15)$&$ 0.788( 3)( 9)$&$ 0.852( 5)(19)$&$ 0.934(12)(23)$&$ 0.953( 9)(19)$ \\
\hline
$\Delta_{51}$& 0.005( 0)( 3)& 0.005( 1)( 2)& 0.007( 1)( 2)& 0.007( 1)( 1)& 0.007( 0)( 2)\\
$\Delta_{52}$& 0.013( 0)( 1)& 0.012( 0)( 1)& 0.009( 1)( 1)& 0.006( 1)( 2)& 0.006( 1)( 2)\\
$\Delta_{53}$& 0.107( 2)( 2)& 0.100( 2)( 3)& 0.081( 2)( 6)& 0.060( 1)( 6)& 0.052( 3)( 6)\\
\hline
\hline
\end{tabular}
\caption{\sl \small Results for the RCs of the PC$^{\,+}$ sector in the $\ri$ 
scheme at the scale $\mu$=2 GeV.}
\label{tab:PC+}
\end{center}
\end{table}
\begin{table}
\begin{center}
\begin{tabular}{|c|c|c|c|c|c|}
\hline
$\beta$&$6.0\,(16^3\times52)$&$6.0\,(24^3\times64)$&$6.2$&$6.4$ & $6.45$ \\
\hline
\hline

\rule[-3mm]{0mm}{8mm}$Z_{11}$&$ 0.622( 3)(10)$&$ 0.614( 3)( 3)$&$ 0.621( 5)( 7)$&$ 0.641( 8)( 8)$&$ 0.642(13)( 9)$ \\
\hline
$\Delta_{12}$&-0.033( 1)( 6)&-0.034( 1)( 5)&-0.029( 1)( 4)&-0.026( 1)( 4)&-0.025( 1)( 3)\\
$\Delta_{13}$& 0.010( 1)( 3)& 0.007( 1)( 1)& 0.005( 1)( 3)& 0.008( 1)( 1)& 0.001( 2)( 6)\\
$\Delta_{14}$&-0.007( 1)( 9)&-0.013( 1)( 6)&-0.016( 2)( 6)&-0.019( 1)( 5)&-0.018( 2)( 5)\\
$\Delta_{15}$&-0.003( 1)( 9)&-0.007( 1)( 6)&-0.012( 2)( 5)&-0.017( 2)( 6)&-0.017( 3)( 6)\\
\hline
\hline

\rule[-3mm]{0mm}{8mm}$Z_{22}$&$ 0.677( 3)( 9)$&$ 0.669( 2)( 3)$&$ 0.700( 4)( 5)$&$ 0.743( 9)(11)$&$ 0.749( 8)( 7)$ \\

\rule[-3mm]{0mm}{8mm}$Z_{23}$&$-0.277( 2)(10)$&$-0.282( 2)( 5)$&$-0.306( 4)( 9)$&$-0.335( 8)(12)$&$-0.340( 5)(10)$ \\
\hline
$\Delta_{21}$&-0.045( 1)( 3)&-0.050( 2)( 3)&-0.040( 2)( 3)&-0.033( 1)( 1)&-0.028( 2)( 4)\\
$\Delta_{24}$& 0.138( 2)( 6)& 0.134( 3)( 1)& 0.113( 3)( 5)& 0.078( 2)( 8)& 0.064( 5)(13)\\
$\Delta_{25}$& 0.012( 1)( 3)& 0.012( 1)( 1)& 0.010( 1)( 2)& 0.007( 1)( 4)& 0.005( 1)( 5)\\
\hline
\hline

\rule[-3mm]{0mm}{8mm}$Z_{32}$&$-0.027( 1)( 1)$&$-0.028( 1)( 1)$&$-0.023( 1)( 1)$&$-0.021( 2)( 1)$&$-0.021( 1)( 2)$ \\

\rule[-3mm]{0mm}{8mm}$Z_{33}$&$ 0.402( 3)(30)$&$ 0.386( 2)(14)$&$ 0.366( 4)(19)$&$ 0.349( 5)(13)$&$ 0.335( 7)(19)$ \\
\hline
$\Delta_{31}$&-0.008( 1)( 5)&-0.011( 1)( 1)&-0.007( 1)( 2)&-0.003( 1)( 3)&-0.004( 2)( 5)\\
$\Delta_{34}$& 0.203( 3)( 9)& 0.195( 3)( 3)& 0.163( 4)( 9)& 0.115( 3)(13)& 0.094( 8)(18)\\
$\Delta_{35}$& 0.000( 0)( 6)& 0.000( 1)( 2)&-0.002( 1)( 2)&-0.002( 1)( 4)&-0.005( 2)( 9)\\
\hline
\hline

\rule[-3mm]{0mm}{8mm}$Z_{44}$&$ 0.309( 5)(44)$&$ 0.300( 3)(15)$&$ 0.280( 5)(20)$&$ 0.258( 5)(16)$&$ 0.238( 7)(37)$ \\

\rule[-3mm]{0mm}{8mm}$Z_{45}$&$-0.007( 1)( 3)$&$-0.007( 1)( 1)$&$-0.016( 1)( 2)$&$-0.020( 2)( 1)$&$-0.019( 4)(18)$ \\
\hline
$\Delta_{41}$& 0.004( 2)(32)& 0.000( 1)( 1)& 0.002( 1)( 4)& 0.003( 2)( 4)&-0.021(14)(63)\\
$\Delta_{42}$&-0.010( 1)( 7)&-0.012( 1)( 2)&-0.004( 1)( 4)& 0.000( 2)(13)&-0.001(10)(55)\\
$\Delta_{43}$& 0.331(18)(85)& 0.303( 5)( 6)& 0.248( 9)( 5)& 0.174( 4)(19)& 0.165(22)(184)\\
\hline
\hline

\rule[-3mm]{0mm}{8mm}$Z_{54}$&$ 0.120( 2)( 3)$&$ 0.124( 2)( 1)$&$ 0.133( 3)( 3)$&$ 0.139( 5)( 2)$&$ 0.133( 7)(15)$ \\

\rule[-3mm]{0mm}{8mm}$Z_{55}$&$ 0.713( 3)(10)$&$ 0.712( 3)( 5)$&$ 0.745( 4)(16)$&$ 0.800(12)(18)$&$ 0.804(13)(26)$ \\
\hline
$\Delta_{51}$&-0.003( 1)( 6)&-0.005( 1)( 4)&-0.010( 1)( 3)&-0.014( 1)( 4)&-0.014( 2)( 4)\\
$\Delta_{52}$& 0.013( 1)( 1)& 0.013( 1)( 1)& 0.009( 1)( 2)& 0.007( 1)( 2)& 0.006( 1)( 4)\\
$\Delta_{53}$&-0.113( 3)( 5)&-0.107( 2)( 2)&-0.087( 3)( 5)&-0.064( 1)( 6)&-0.058( 2)( 6)\\
\hline
\hline
\end{tabular}
\caption{\sl \small Results for the RCs of the PC$^{\,-}$ sector in the $\ri$ 
scheme at the scale $\mu$=2 GeV.}
\label{tab:PC-}
\end{center}
\end{table}
\begin{table}
\begin{center}
\begin{tabular}{|c|c|c|c|c|c|}
\hline
$\beta$&$6.0\,(16^3\times52)$&$6.0\,(24^3\times64)$&$6.2$&$6.4$ & $6.45$ \\
\hline
\hline

\rule[-3mm]{0mm}{8mm}$Z_{11}$&$ 0.608( 4)(14)$&$ 0.604( 3)( 5)$&$ 0.635( 5)( 4)$&$ 0.671( 9)( 8)$&$ 0.681( 7)( 3)$ \\
\hline
\hline

\rule[-3mm]{0mm}{8mm}$Z_{22}$&$ 0.673( 3)(10)$&$ 0.666( 3)( 3)$&$ 0.697( 4)( 3)$&$ 0.738( 8)(11)$&$ 0.744( 8)( 9)$ \\

\rule[-3mm]{0mm}{8mm}$Z_{23}$&$ 0.279( 2)( 9)$&$ 0.281( 2)( 3)$&$ 0.307( 4)( 9)$&$ 0.334( 8)(11)$&$ 0.335( 6)(10)$ \\
\hline
\hline

\rule[-3mm]{0mm}{8mm}$Z_{32}$&$ 0.028( 1)( 1)$&$ 0.028( 1)( 1)$&$ 0.024( 1)( 1)$&$ 0.023( 2)( 1)$&$ 0.021( 1)( 2)$ \\

\rule[-3mm]{0mm}{8mm}$Z_{33}$&$ 0.392( 3)(28)$&$ 0.375( 3)(14)$&$ 0.353( 4)(21)$&$ 0.338( 4)(14)$&$ 0.331( 6)(17)$ \\
\hline
\hline

\rule[-3mm]{0mm}{8mm}$Z_{44}$&$ 0.476( 3)(30)$&$ 0.457( 3)(15)$&$ 0.437( 4)(22)$&$ 0.425( 5)(16)$&$ 0.417( 5)(22)$ \\

\rule[-3mm]{0mm}{8mm}$Z_{45}$&$-0.024( 1)( 1)$&$-0.022( 1)( 1)$&$-0.019( 1)( 1)$&$-0.016( 2)( 1)$&$-0.015( 2)( 3)$ \\
\hline
\hline

\rule[-3mm]{0mm}{8mm}$Z_{54}$&$-0.255( 2)(11)$&$-0.260( 2)( 5)$&$-0.295( 3)(12)$&$-0.333( 6)(14)$&$-0.339( 6)(13)$ \\

\rule[-3mm]{0mm}{8mm}$Z_{55}$&$ 0.787( 4)(15)$&$ 0.790( 3)( 9)$&$ 0.854( 5)(19)$&$ 0.936(12)(24)$&$ 0.952(10)(19)$ \\
\hline
\hline
\end{tabular}
\caption{\sl \small Results for the RCs of the PV$^{\,+}$ sector in the $\ri$ 
scheme at the scale $\mu$=2 GeV.}
\label{tab:PV+}
\end{center}
\end{table}
\begin{table}
\begin{center}
\begin{tabular}{|c|c|c|c|c|c|}
\hline
$\beta$&$6.0\,(16^3\times52)$&$6.0\,(24^3\times64)$&$6.2$&$6.4$ & $6.45$ \\
\hline
\hline

\rule[-3mm]{0mm}{8mm}$Z_{11}$&$ 0.623( 3)(10)$&$ 0.614( 3)( 4)$&$ 0.621( 5)( 5)$&$ 0.642( 8)( 8)$&$ 0.642(11)( 8)$ \\
\hline
\hline

\rule[-3mm]{0mm}{8mm}$Z_{22}$&$ 0.673( 3)(10)$&$ 0.666( 3)( 4)$&$ 0.696( 4)( 3)$&$ 0.737( 8)(11)$&$ 0.744( 8)( 9)$ \\

\rule[-3mm]{0mm}{8mm}$Z_{23}$&$-0.279( 2)( 9)$&$-0.281( 2)( 3)$&$-0.307( 4)( 9)$&$-0.333( 8)(12)$&$-0.335( 6)(10)$ \\
\hline
\hline

\rule[-3mm]{0mm}{8mm}$Z_{32}$&$-0.028( 1)( 1)$&$-0.028( 1)( 1)$&$-0.024( 1)( 1)$&$-0.023( 2)( 1)$&$-0.021( 1)( 2)$ \\

\rule[-3mm]{0mm}{8mm}$Z_{33}$&$ 0.392( 3)(28)$&$ 0.375( 3)(14)$&$ 0.353( 4)(21)$&$ 0.338( 4)(14)$&$ 0.331( 6)(17)$ \\
\hline
\hline

\rule[-3mm]{0mm}{8mm}$Z_{44}$&$ 0.312( 3)(36)$&$ 0.295( 3)(16)$&$ 0.271( 4)(22)$&$ 0.249( 4)(21)$&$ 0.218(15)(80)$ \\

\rule[-3mm]{0mm}{8mm}$Z_{45}$&$-0.006( 1)( 3)$&$-0.007( 1)( 1)$&$-0.016( 1)( 3)$&$-0.020( 2)( 1)$&$-0.020( 5)(52)$ \\
\hline
\hline

\rule[-3mm]{0mm}{8mm}$Z_{54}$&$ 0.126( 1)( 4)$&$ 0.126( 2)( 1)$&$ 0.135( 3)( 3)$&$ 0.140( 5)( 1)$&$ 0.130(10)(36)$ \\

\rule[-3mm]{0mm}{8mm}$Z_{55}$&$ 0.715( 3)(12)$&$ 0.714( 3)( 7)$&$ 0.746( 4)(17)$&$ 0.802(12)(18)$&$ 0.806(16)(32)$ \\
\hline
\hline
\end{tabular}
\caption{\sl \small Results for the RCs of the PV$^{\,-}$ sector in the $\ri$ 
scheme at the scale $\mu$=2 GeV.}
\label{tab:PV-}
\end{center}
\end{table}

\section*{Conclusions}
In this paper we have presented the results of an extensive lattice calculation
of the RCs of bilinear and four-quark $\Delta F=2$ and $\Delta I=3/2$ operators
by using the $\ri$ NPR method. Several sources of systematic errors, including
discretization errors and finite volume effects, have been investigated. When 
possible, we have also compared the results with those obtained by using other 
non-perturbative approaches, like the WI and the SF methods. In these cases, we 
find an agreement which is typically at the level of 1\%. This comparison 
supports the conclusion that the $\ri$ method allows to obtain an accurate 
non-perturbative determination of the RCs of lattice operators. At the same 
time, the method is extremely simple to implement and can be applied to a large 
class of operators. The only notable exception is represented by those operators
the renormalization of which requires subtraction of power divergences. In the 
latter case, and in particular in the case of the $\Delta I=1/2$ four-fermion 
operators, the use of non gauge invariant correlation functions renders the 
applicability of the $\ri$ method extremely difficult, if not impossible in 
practice. In order to renormalize these operators the gauge invariant NPR method
in $x$-space~\cite{XS}-\cite{lat03_rey} might be the method of choice.

\section*{Acknowledgments}
It is a pleasure to thank C.J.D.~Lin, A. Le Yaouanc, C.T.~Sachrajda and M. 
Testa for many interesting discussions on the subject of this paper.

\section*{Appendix A}
In this appendix we show that the RCs of bilinear quark operators obtained with 
the $\ri$ method in the chiral limit, at sufficiently large values of the 
external momentum and at zero momentum transfer, are automatically improved at
$\Oa$.

The renormalized and improved version of the bilinear operator $O_\Gamma=\qbar 
\Gamma q$ has the form~\cite{impos}: 
\beq
\hat O_\Gamma^I = Z_\Gamma O_\Gamma^I = Z_\Gamma \left( \,\qbar \Gamma q \,+ \,
a \,c_\Gamma O_{4,\Gamma}\, + \, a \,\cp_\Gamma E_\Gamma  \right)
\label{eq:Ogamma}
\eeq
where $Z_\Gamma$ is the multiplicative RC in which we also include here the 
$\Oa$ linear dependence on the quark mass, $Z_\Gamma = Z^0_\Gamma \left(1 + 
b_\Gamma a m\right)$. The coefficients $b_\Gamma$, $c_\Gamma$, $\cp_\Gamma$ 
express the mixing of $O_\Gamma$ with higher-dimension operators. This mixing 
has to be taken into account in order to improve the operator at $\Oa$. The 
operator $E_\Gamma$ in eq.~(\ref{eq:Ogamma}) is given by:
\beq
E_\Gamma = \qbar \left[ \Gamma (\rDslash + m_0) + ( - \lDslash + m_0) \Gamma 
\right] q 
\label{eq:Egamma}
\eeq
where $\rDslash + m_0$ is a shorthand for the entire lattice fermion operator, 
including the Wilson and the SW-clover terms. Therefore $E_\Gamma$ vanishes 
by the equation of motion and it only contributes to contact terms when inserted
in correlation functions. The operator $O_{4,\Gamma}$ in eq.~(\ref{eq:Ogamma}) 
is a gauge-invariant, dimension-four operator which does not vanish by the 
equation of motion. In the cases of the vector, axial-vector and tensor
operators, $O_{4,\Gamma}$ is given by
\bea
\label{eq:o4gamma}
O_{4,V} &=& \partial_\nu \, (\qbar \sigma_{\mu,\nu} q) \nonumber \\
O_{4,A} &=& \partial_\mu \, (\qbar \gamma^5 q) \\
O_{4,T} &=& \partial_\mu \, (\qbar \gamma_{\nu} q) - 
\partial_\nu \, (\qbar \gamma_{\mu} q) \nonumber
\eea
while this mixing is absent for the scalar and pseudoscalar densities.

In order to improve the off-shell correlation functions, we also consider the 
renormalized improved quark and antiquark fields~\cite{impos},
\bea
\label{eq:qhat}
\hat q &=& Z_q^{-1/2} \left[
1 +a \,\cp_q(\rDslash + m_0) + a \,\cngi \rdslash \right]q \\ \nonumber
\hat \qbar &=& Z_q^{-1/2} \, \qbar \left[
1 +a \,\cp_q(- \lDslash + m_0) - a \,\cngi \ldslash \right] 
\eea
with $Z_q = Z^0_q \left(1 + b_q a m\right)$. In terms of these fields we can
define the renormalized improved quark propagator: 
\beq
\hat{S}(p) = \int d^4x \, e^{-i p\cdot x} \, \langle \hat q(x) \hat\qbar(0) 
\rangle \, ,
\eeq
which, by using eq.~(\ref{eq:qhat}), can be related to the corresponding lattice
quantity $S(p)$ by
\beq
\hat{S}(p) = Z_q^{-1} \, \left[ S(p) + 2 a \,\cp_q + 2 a \,\cngi \, i \pslash\, 
S(p) \right] \,
\eeq
up to $\Oaa$ terms.

Let us now discuss the effect of the off-shell improvement, defined by 
eqs.~(\ref{eq:Ogamma}) and (\ref{eq:qhat}), in the determination of the RCs with
the $\ri$ method. The relevant renormalized improved correlation function is 
defined in terms of improved operators and external fields:
\bea
&& {G}_\Gamma^I(p,p') = \int d^4x \, d^4y \, e^{-i p\cdot x + i p'\cdot y} \, 
\langle \hat q(x) O_\Gamma^I(0) \hat\qbar(y) \rangle = \nonumber \\
&& \qquad \int d^4x \, d^4y \, e^{-i p\cdot x + i p'\cdot y} \, \langle 
\hat q(x) \left( \qbar \Gamma q \,+ \, a \,c_\Gamma O_{4,\Gamma}\, + \, a \,
\cp_\Gamma E_\Gamma \right)_0 \hat\qbar(y) \rangle \, .
\label{eq:gamp}
\eea
The $\ri$ method consists in imposing that the forward amputated Green function,
\beq
\Lambda_\Gamma^I(p) = \hat{S}(p)^{-1} {G}_\Gamma^I(p,p) \hat{S}(p)^{-1} \, ,
\label{eq:lamp}
\eeq
computed in a fixed gauge and renormalized at a given scale $p^2=\mu^2$, is 
equal to its tree-level value. In practice, this condition is implemented by 
requiring:
\beq
Z_\Gamma \, \Gamma_\Gamma^I (p) \vert_{p^2=\mu^2} = Z_\Gamma \, 
\Tr [\Lambda_\Gamma^I (p)\, P_\Gamma] \vert_{p^2=\mu^2} =1
\label{eq:ri}
\eeq
where $P_\Gamma$ is the Dirac projector.

The Green functions $G_\Gamma^I(p)$ and $\Lambda_\Gamma^I(p)$ can be easily
related to their unimproved counterparts. By using 
eqs.~(\ref{eq:Ogamma})-(\ref{eq:gamp}), and up to terms of $\Oaa$, one finds:
\bea 
G_\Gamma^I(p) &=& G_\Gamma(p) \,+ \, a \,c_\Gamma \, G_{4,\Gamma}(p) \, + \, 
a\, (\cp_\Gamma + \cp_q) \left( \Gamma \, \hat S(p) + \hat S(p) \Gamma \right)  
+ \nonumber \\ && 
a \, \cngi \left( i\pslash \, G_\Gamma^I(p) + G_\Gamma^I(p) \, i\pslash \right)
\label{eq:gimp}
\eea
where $G_\Gamma(p)$ is the Green function of the unimproved operator 
$O_\Gamma$, constructed with unimproved external quark fields, and 
$G_{4,\Gamma}(p)$ is the analogous quantity for the operator $O_{4,\Gamma}$ 
defined in eq.~(\ref{eq:Ogamma}). The improved function $\Lambda_\Gamma^I(p)$ 
is then obtained from $G_\Gamma^I(p)$ by amputating the external legs with 
improved quark propagators, according to eq.~(\ref{eq:lamp}). The result is: 
\bea 
&& \Lambda_\Gamma^I(p) = \Lambda_\Gamma(p) \,+ \, a \,c_\Gamma \, 
\Lambda_{4,\Gamma}(p) \, + \, a\, (\cp_\Gamma + \cp_q) \left( \Gamma \, 
\hat S(p)^{-1} + \hat S(p)^{-1} \Gamma \right)  - \nonumber \\
&& \qquad 2 a \, \cp_q \, Z_q^{-1} \left( \hat S(p)^{-1} \Lambda_\Gamma^I(p) + 
\Lambda_\Gamma^I(p) \hat S(p)^{-1} \right) -
a \, \cngi \left( i \pslash \, \Lambda_\Gamma^I(p) + \Lambda_\Gamma^I(p) \, 
i \pslash \right)
\label{eq:lambda}
\eea
Finally, by projecting eq.~(\ref{eq:lambda}) onto the tree-level form factor 
with the projector $P_\Gamma$, one obtains:
\bea 
&& \Gamma_\Gamma^I(p) = \Gamma_\Gamma(p) \, + \, a \,c_\Gamma \, 
\Gamma_{4,\Gamma}(p) \, + \, 2\, a\, (\cp_\Gamma + \cp_q) \Tr [\hat S(p)^{-1}]
- \nonumber \\
&& \qquad 2 a \, \cp_q \, Z_q^{-1} 
\Tr \left( \hat S(p)^{-1}  \, \Lambda_\Gamma^I(p) \, P_\Gamma -
\Lambda_\Gamma^I(p) \, \hat S(p)^{-1} \, P_\Gamma \right)  + \nonumber \\
&& \qquad a \, \cngi \Tr \left( i \pslash  \, \Lambda_\Gamma^I(p) \, P_\Gamma + 
\Lambda_\Gamma^I(p) \, i \pslash \, P_\Gamma \right) \, .
\label{eq:gamma}
\eea
The various contributions to the amputated projected Green function in 
eq.~(\ref{eq:gamma}) are easily identified. The third term on the r.h.s. of 
eq.~(\ref{eq:gamma}) comes from the operators vanishing by the equation of 
motion and it is proportional to the trace of the improved inverse quark 
propagator. Since at large values of $p^2$ this quantity is proportional to 
the quark mass, such a term does not affect the determination of the 
renormalization constants in the chiral limit.

One may be easily convinced that also the last two terms in eq.~(\ref{eq:gamma})
proportional to $\cp_q$ and $\cngi$ respectively, vanish in this limit. We first
observe that, for vanishing quark mass, the improved propagator $\hat S(p)^{-1}$
at large $p^2$ is proportional to $\pslash$. Therefore, the two terms have the 
same form. In addition, we note that contributions from $\Lambda_\Gamma^I(p)$ 
proportional to the tree-level form factor, $\Gamma$, vanish because they reduce
to $\Tr (\pslash)$. For the same reason, one also finds the vanishing of all 
possible contributions coming from different (dimensionless) form factor which 
do not depend on the quark mass. For instance, in the case of the vector 
current, the Green function $\Lambda_\mu^I(p)$ also contains a term proportional
to $p_\mu \pslash/p^2$, which again gives a contribution vanishing as $\Tr 
(\pslash)$ to the last two terms of eq.~(\ref{eq:gamma}). A non vanishing 
contribution may come from a form factor proportional to $m \pslash \gamma_\mu/
p^2$, but then this contribution vanishes in the chiral limit.

In the case of the vector, axial-vector and tensor operators, on-shell 
$\Oa$-improvement also requires to consider the mixing of the bare operators 
$O_\Gamma$ with the dimension-4 operators $O_{4,\Gamma}$, whose contribution is 
represented by the second term on the r.h.s. of eq.~(\ref{eq:gamma}). For the 
scalar and pseudoscalar densities this mixing is absent. In the other cases, 
the operators $O_{4,\Gamma}$ are listed in eq.~(\ref{eq:o4gamma}). A simple
observation is that the operators $O_{4,\Gamma}$ all have the form of a 
4-divergence. Therefore, their forward correlation functions, $\Gamma_{4,\Gamma}
(p)$ in eq.~(\ref{eq:gamma}), vanish identically. This completes the proof that 
only the unimproved correlator $\Gamma_\Gamma(p)$ survives at large $p^2$ and in
the chiral limit on the r.h.s. of eq.~(\ref{eq:gamma}) and contributes to the 
calculation of the RCs with the $\ri$ method. Such a calculation can thus be 
performed even in the lacking of a non-perturbative determination of the 
coefficients $c_\Gamma$ and $\cp_\Gamma$. Note also that, although the 
correlation function $\Gamma_\Gamma^I(p)$ is obtained from $G_\Gamma^I(p,p)$ by 
amputating the external legs with improved quark propagators, in practice, in 
the chiral limit, improving the quark propagator is also an unnecessary step 
for the determination of the RCs.

\section*{Appendix B}
In this appendix we show that, in the case of four-fermion operators, both 
single and double Goldstone boson poles can appear in the chiral limit in the 
relevant correlation functions, and may thus affect the $\ri$ NPR procedure at 
finite values of the external momenta. For convenience, we work with non 
vanishing quark masses and we will consider the zero quark mass limit only at 
the end of the calculation. 

Let us consider the Green function of a four fermion operator ${\cal O}_i$ with
four external quark states and two different external momenta
\bea
G_i(p,p')&=&\int d^4x_1\, d^4x_2\, d^4x_3\, d^4x_4\, 
e^{-ip(x_1+x_3)+ip'(x_2+x_4)} \, \times \nn\\
&& \langle 0|T[\psi^a(x_1)\, \bar\psi^b(x_2)\, {\cal O}_i(0)\, \psi^c(x_3)\,
\bar\psi^d(x_4)]|0\rangle \nn \\
&\equiv& 
\langle 0|T[\tilde{\psi^a}(p)\, \tilde{\bar\psi^b}(p')\, {\cal O}_i(0)\, 
\tilde{\psi^c}(p)\, \tilde{\bar\psi^d}(p')]|0\rangle 
\label{eq:gaxial1}
\eea
By applying the LSZ reduction formula we can write
\bea
G_i(p,p')&=&\langle 0|T[\psi^a(0)\,\tilde{\bar \psi^b}(p')]|\pi^{ba}(q)\rangle\,
\frac{i}{q^2-m_\pi^2}\,
\langle\pi^{ba}(q)|T[{\cal O}_i(0)\,\tilde{\psi^c}(p)\,\tilde{\bar\psi^d}(p')]
|0\rangle\, +  \nn\\
&& NP\left[\langle 0|T[\tilde{\psi^a}(p)\, \tilde{\bar\psi^b}(p')\, {\cal O}_i
(0)\, \tilde{\psi^c}(p)\, \tilde{\bar\psi^d}(p')]|0\rangle\right]\,+\,\ldots
\label{eq:lszaxial1}
\eea
where $q_\mu=p_\mu-p'_\mu$ and we have identified the Goldstone boson with the
$\pi$ meson. In eq.~(\ref{eq:lszaxial1}) the symbol $NP[\dots]$ indicates the 
part of the correlation function without pions in the intermediate states and 
the dots represent terms with pion propagators corresponding to pions different
from $|\pi^{ba}(q)\rangle$ (i.e. $|\pi^{da}(q)\rangle$, $|\pi^{cd}(q)\rangle$ 
and $|\pi^{bc}(q)\rangle$). If we now apply the reduction formula again we 
obtain
\bea
\label{eq:lszaxial2}
&& G_i(p,p')=\langle0|T[\psi^a(0)\tilde{\bar\psi^b}(p')]|\pi^{ba}(q)\rangle
\frac{i}{q^2-m_\pi^2}\, \times \nn\\
&&\qquad \left\{ \langle\pi^{ba}(q)|{\cal O}_i(0)|\pi^{cd}(q)\rangle \frac{i}
{q^2-m_\pi^2} \langle\pi^{cd}(q)|T[\psi^c(0) \tilde{\bar\psi^d}(p')]|0\rangle \,
+ \right. \nn\\
&&\qquad \left. NP\left[\langle\pi^{ba}(q)|T[{\cal O}_i(0)\tilde
\psi^c(p) \tilde{\bar\psi^d}(p')]|0\rangle\right]\right\} +\ldots\,.
\eea

In the implementation of the $\ri$ method we have considered the case $p_\mu=
p'_\mu$, corresponding to $q_\mu=0$. From eq.~(\ref{eq:lszaxial2}) it is 
then clear that both single and double pion poles can appear in the amputated 
Green functions in the chiral limit. In the case of parity violating operators, 
however, since by parity $\langle\pi^{ba}(0)|{\cal O}_i^{PV}(0)|\pi^{cd}(0)
\rangle=0$, only single poles can be present.

\end{document}